\documentclass[aip,pop,reprint]{revtex4-2}
\usepackage{graphicx}
\usepackage{amsmath}
\usepackage{mathrsfs}
\usepackage{array}
\usepackage{siunitx}
\usepackage[breaklinks=true,colorlinks=true,linkcolor=blue,urlcolor=blue,citecolor=blue]{hyperref}
\graphicspath{{./Figures/}}

\begin{document}

\title[Formation and Study of a Spherical Plasma Liner for Plasma-Jet-Driven Magneto-Inertial Fusion]{Formation and Study of a Spherical Plasma Liner for Plasma-Jet-Driven Magneto-Inertial Fusion}
\author{A. L. LaJoie}
 \affiliation{University of New Mexico}
 
\author{F. Chu}%
\affiliation{
Los Alamos National Laboratory
}%

\author{A. Brown}%
\affiliation{
Los Alamos National Laboratory
}%

\author{S. Langendorf}
\affiliation{
Los Alamos National Laboratory
}%

\author{J. P. Dunn}
\affiliation{
Los Alamos National Laboratory
}%

\author{G. A. Wurden}
\affiliation{
Los Alamos National Laboratory
}%

\author{F. D. Witherspoon}
\affiliation{%
HyperJet Fusion Corp.
}%

\author{A. Case}
\affiliation{%
HyperJet Fusion Corp.
}%

\author{M. Luna}
\affiliation{%
HyperJet Fusion Corp.
}%

\author{J. Cassibry}
\affiliation{%
University of Alabama at Huntsville
}%

\author{A. Vyas}
\affiliation{%
University of Alabama at Huntsville
}%

\author{M. Gilmore}
\affiliation{%
University of New Mexico
}%
\date{\today}

\begin{abstract}
Plasma-jet-driven magneto-inertial fusion (PJMIF) is an alternative approach to controlled nuclear fusion which aims to utilize a line-replaceable dense plasma liner as a repetitive spherical compression driver. In this experiment, first measurements of the formation of a spherical Argon plasma liner formed from 36 discrete pulsed plasma jets are obtained on the Plasma Liner Experiment (PLX). Properties including liner uniformity and morphology, plasma density, temperature, and ram pressure are assessed as a function of time throughout the implosion process and indicate an apparent transition from initial kinetic inter-jet interpenetration to collisional regime near stagnation times, in accordance with theoretical expectation. A lack of primary shock structures between adjacent jets during flight implies that arbitrarily smooth liners may be formed by way of corresponding improvements in jet parameters and control. The measurements facilitate the benchmarking of computational models and understanding the scaling of plasma liners towards fusion-relevant energy density.
\end{abstract}

\maketitle

\section{\label{sec:level1}Introduction\protect\\}
Magneto-inertial fusion (MIF, also called magnetized target fusion or MTF) is an approach to controlled nuclear fusion which joins aspects of both magnetic confinement fusion and inertial confinement fusion methods, enabling controlled fusion at more modest values of both confinement time and density \cite{myID_577,myID_578,myID_579,myID_563,myID_520} than those in the purely inertial or magnetic confinement counterparts. Many MIF compression approaches and configurations have been proposed and studied, including the Magnetized Liner Inertial Fusion (MagLIF) Z-pinch configuration \cite{myID_566}, shear-flow-stabilized Z pinches \cite{myID_567}, spherical piston-driven liquid metal liners \cite{myID_572}, rotationally stabilized liquid metal liners\cite{myID_573}, and staged-gas-puff Z-pinches\cite{myID_574}. Notably, MagLIF experiments have successfully produced thermonuclear conditions and significant neutron yields from the compression of preheated magnetized deuterium fuel, on implosion timescales significantly longer than can be tolerated in pure inertial fusion.

The plasma-jet-driven magneto-inertial fusion (PJMIF) concept is an MIF approach in which the requisite heavy compressive liner is formed from the merging and spherical convergence of an array of discrete plasma jets. PJMIF aims to access an attractive subsection of the MIF physics parameters space, namely that of the spherical, moderate velocity (50-150 km/s) compression of high-beta high-Hall-parameter targets which have shown promise in MagLIF \cite{myID_545, myID_519,myID_520}. If successful, this architecture would enable repetitive line-replaceable operation with a significant standoff distance between the fusion plasma and driver hardware. Achieving economical repetitive operation is a significant challenge facing all proposed pulsed fusion concepts, in particular those seeking to utilize electrical pulsed-power as the primary energy source \cite{myID_563,myID_545}. 

There remain several key questions regarding the viability of the PJMIF approach, including whether conditions of sufficient energy density and fusion gain can be attained while maintaining the physical standoff that constitutes a key appeal of the approach, whether sufficient uniformity of a spherical liner can be attained from the merger of discrete jets in the face of jet-on-jet shock waves and Rayleigh-Taylor instabilities \cite{myID_534,myID_522,myID_521}, and whether effective magnetic insulation of a standoff target plasma can be achieved in spherical implosion geometry \cite{myID_565}. Prior investigations have shed light on several of these questions, with analytical and computational modeling predicting the feasibility of significant fusion gain \cite{myID_575}, the realization of high-performance modular plasma guns to produce controlled plasma jets \cite{myID_513}, and laboratory experiments studying the merging and formation processes of multiple plasma jets at representative velocities \cite{myID_516,myID_570,myID_549,myID_512, myID_576}. Much of the experimental work has been done at the Plasma Liner Experiment (PLX) at Los Alamos, a moderate-scale experimental effort towards understanding the challenges and scaling of PJMIF. Early experimental campaigns at PLX focused on plasma gun development and generating suitable pulsed-power-driven jets \cite{myID_513, myID_580}, while more recent campaigns have been aimed at understanding and characterizing jet merging behavior in obliquely merging jet pairs \cite{myID_516,myID_570}, and liner subsection formation of up to 6 or 7 jets \cite{myID_563,myID_549}.

In the present work, fully spherical plasma liners have been formed and investigated at the Plasma Liner Experiment (PLX). This experiment uses 36 individual argon plasma jets, enabling observation of the integrated liner assembly process for the first time. The primary aim is to assess the uniformity and plasma characteristics of the full spherical liner formed by discrete jets and determine where future work must focus to close the gap between present liner quality and fusion-relevant characteristics. The discussion of this full liner experimental campaign is organized as follows: Section \ref{sec:Experiment} details the pulsed power setup and plasma jet formation and tuning, Section \ref{sec:Diagnostics} describes what plasma properties are measured and the diagnostics used to do so, and Section \ref{sec:Results} presents the results of the diagnostic suite and implications for PJMIF as a whole. Considerations necessary for continuation and an eventual reactor-scale device are explored in Section \ref{sec:Conclusions} as well as a concluding summary of points.

\section{\label{sec:Experiment}Experiment\protect\\}

\begin{figure}
\includegraphics[width=8.5cm]{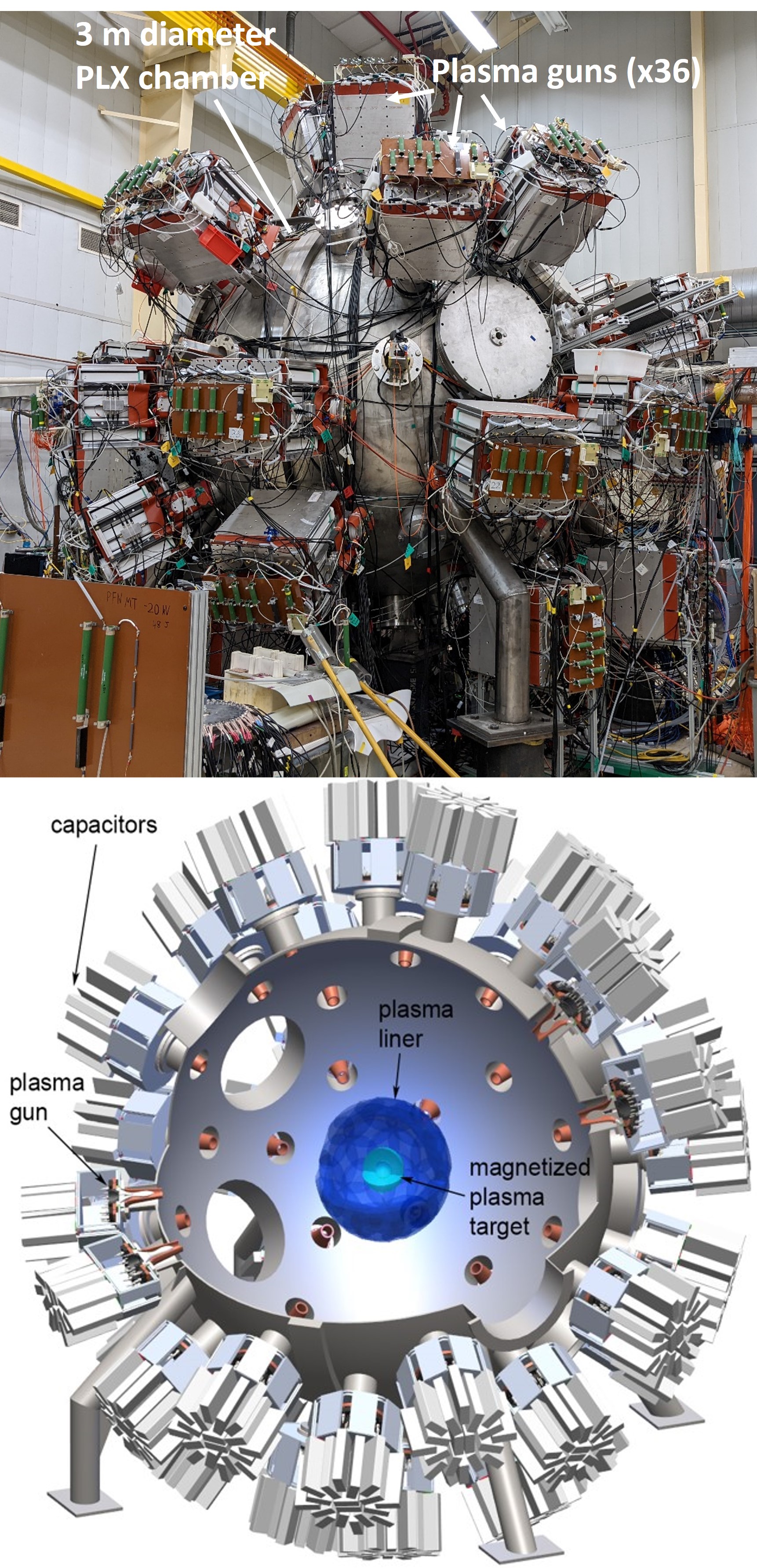}
\caption{\label{fig:PLX-Photo_and_render2} Photo of the PLX chamber alongside a rendering of the fully assembled plasma liner and target. The present study focuses exclusively on the plasma liner behavior.}
\end{figure}

The PLX device consists of a 3-meter diameter chamber and 36 attached plasma guns, shown in Fig. \ref{fig:PLX-Photo_and_render2} along with a rendered cross section of the chamber and notional plasma liner and target within \cite{myID_519}. The plasma guns are positioned quasi-spherically over the chamber, with an average full-angle between adjacent guns of 36 degrees and a minimum full-angle of 22 degrees due to the relative positioning of available gun attachment ports. Chamber vacuum base pressure is typically around 1 $\mu$torr before plasma jets are injected into the chamber, increasing to $\sim$ 10 $\mu$torr when a turbopump gate valve is closed before the shot.

The plasma guns that supply the liner-forming plasma jets are contoured-gap coaxial plasma railguns - the contoured shape of the inner electrode alleviates the blowby instability in such geometries that can hinder outgoing jet quality. The guns used in the present PLX work are similar to those described in detail in Ref[\onlinecite{myID_513}], although the current PLX ones are a more compact design. The stored energy per gun per pulse is 4.5-6 kJ, which accelerates argon plasma jets of total mass $\sim$1 mg to typical tunable velocities between 30 and 70 km/s out of a barrel that is 8.5 cm in diameter. Most of the plasma gun capacitor banks and spark-gap switches are mounted directly onto the chamber to minimize transfer inductance to the plasma load and increase energy coupling. Due to space access limitations around the lowest ports of the PLX chamber, nine guns are instead linked to their capacitor banks via an array of transmission lines, referred to hereafter as transmission line (TL) guns. Plasma properties of the jets at the time of ejection have been explored in previous publications~\cite{myID_570}; typically, for a given jet around ejection time, $T_e \approx 1.5-2.0$ eV and $n_e \approx 10^{16}$ cm\textsuperscript{-3}.

Four distinct collections of capacitor banks are activated when firing a PLX shot. The first is a high voltage (19.0 kV) modified-Blumlein trigger network for the plasma gun gas valves. This sends 36 trigger pulses down coaxial cables to 36 spark gap switches that operate the gas valve capacitor banks of the plasma guns. Each plasma gun has a separate 100 J capacitor bank that actuates the gun's fast gas puff valve, injecting neutral argon gas injection into the guns.  An adjustable time delay passes while gas valves operate and neutral gas flows into the bore of the plasma guns, typically 480 microseconds in the current experiments.  Thirdly, another modified-Blumlein trigger network fires, to breakdown the parallel spark-gaps for the main gun banks. These banks have 575 $\mu$F rated to 5 kV peak voltage, and provide most of the current to form and accelerate the plasma jet. The main bank is discharged across a set of six spark-gap switches to the coaxial rail electrodes, again to minimize transfer inductance and prolong switch electrode lifetime. Operating voltages are set to 4.0 kV for non-TL guns, while the TL guns described above are charged to 4.5 kV to compensate for the increased transfer inductance of their transmission lines. Time $t=0$ is defined as when the main trigger pulse for the main accelerating banks is sent. Pressure adjustments can be made to many of the spark-gap switches on a per-gun basis to ensure they all break down properly with the trigger pulse. A resistive circuit in parallel with the pulse-forming-network (PFN) spark-gap trigger switches produces pre-ionization at the breech - this preionization reduces time jitter of the bulk plasma breakdown and subsequent acceleration generated by the main bank.

Tuning the individual jet velocity is essential in this experiment, in order to ensure the jets arrive in the chamber at the same time and merge together with as high degree of spherical symmetry as possible. The component voltages described above can be adjusted for varying effects: a higher main trigger voltage can reduce firing jitter but increases probability of prefiring and arcing between components; main bank charge voltage establishes the overall energy with which the jets are ionized and accelerated; GV charge voltages establish how much gas is injected into the breach and subsequently accelerated. The most robust means of tuning jet velocity is adjusting the GV bank voltage and therefore the injected mass into which the accelerating PFN energy is deposited. To achieve a nominal jet front velocity $v_j$ of 50 km/s for each of the 36 jets, adjustments to this GV bank voltage for each individual gun are made approximately 2 seconds to firing to fine-tune the respective jet velocity. This tuning method enables significantly more flexible and easier adjustment than the previous method of introducing ballast impedence wires in series with the GV described in Appendix B of Ref[\onlinecite{myID_549}]. Results in this paper are obtained using a nominal best tune of each gun to 50 km/s. The process of determining the best tune is described in Section \ref{sec:Diagnostics_Photodiodes}. Mach number $M = v_{jet} / c_s$, with $c_s = [\gamma p/\rho]^{1/2} = [\gamma / m_{i}( \bar{Z}T_e +T_i)]^{1/2}$, where adiabatic index $\gamma=5/3$ is assumed and $m_i$ is the Argon ion mass. To estimate $M$, we will use calculations from previous work examining the exiting jet for temperature $T_e = T_i \approx 1.5$ and $\bar{Z} \approx 1.0$. With these values, estimated ejection Mach number in the present experiment is around $M \approx 14$.

\begin{figure}
\includegraphics[width=8.5cm]{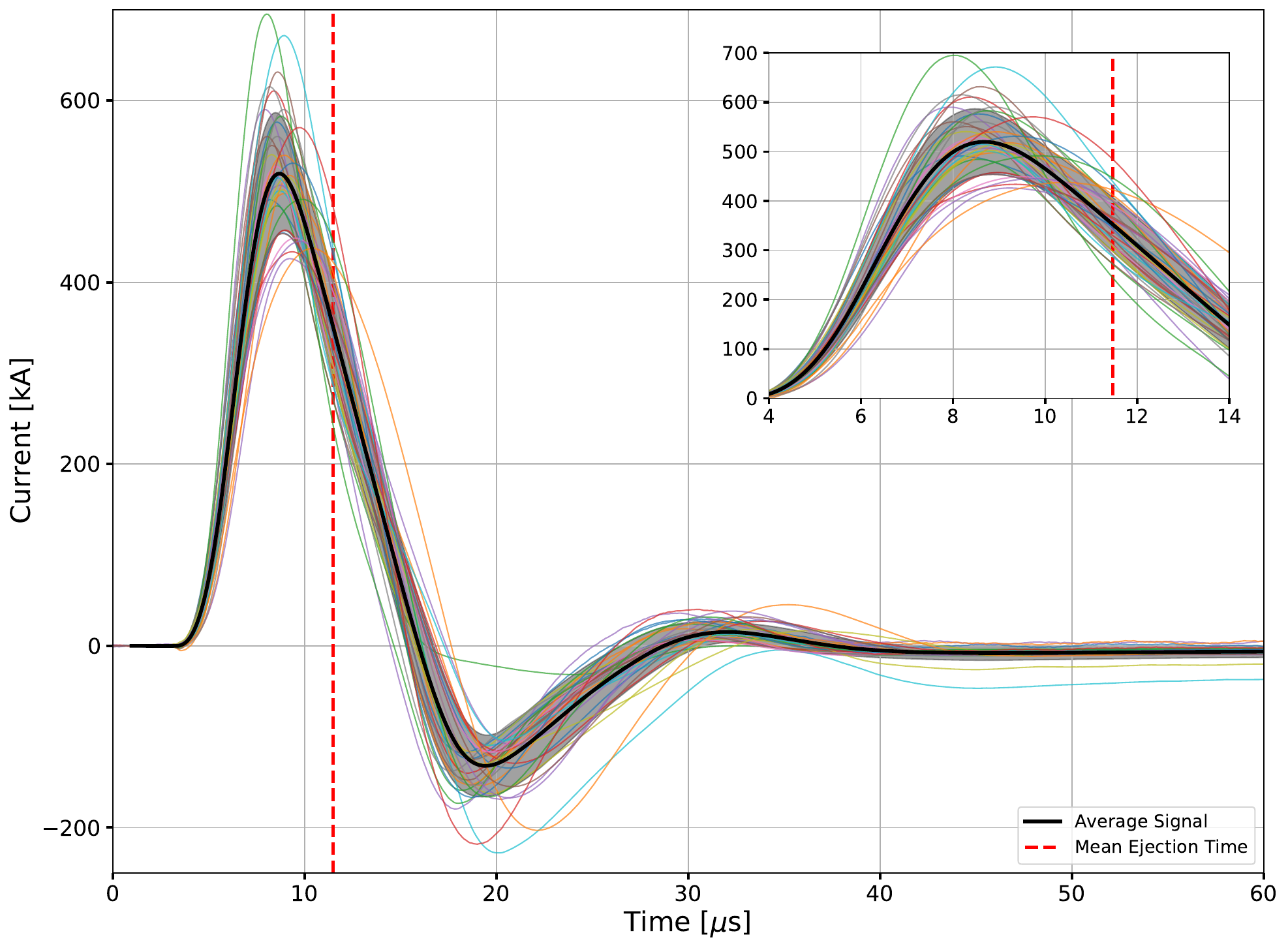}
\caption{\label{fig:AllRogos_OnePlot_Shot2187} Rogowski coil signals for all 36 guns on a shot. Mean signal and standard deviation zone are shown as black curve and grey band. The dashed line is the mean ejection time of all jets. The inset shows a zoomed in portion around the jet ejection time.}
\end{figure}

Main bank and GV discharges are monitored using a set of Rogowski coils on each gun: six on the main bank transmission lines and one on the GV transmission line. Signals from these are sent to a static integrator box to convert the measured $dI/dt$ to $I(t)$ where $I$ is current. It has been empirically determined that jet ejection time corresponds to about 2.5 $\mu$s after the peak of the main bank Rogowski current signal. Figure \ref{fig:AllRogos_OnePlot_Shot2187} shows a set of all 36 main bank Rogowski signals, the mean signal and standard deviation zone around it, and the mean ejection time given by the dashed line. Current ringing is observed after the primary accelerating current corresponding to the signal peak, as the underdamped capacitor circuit discharges its remaining energy into residual trailing plasma in the gun. 

\section{\label{sec:Diagnostics}Diagnostics\protect\\}
\begin{figure}
\includegraphics[width=8.5cm]{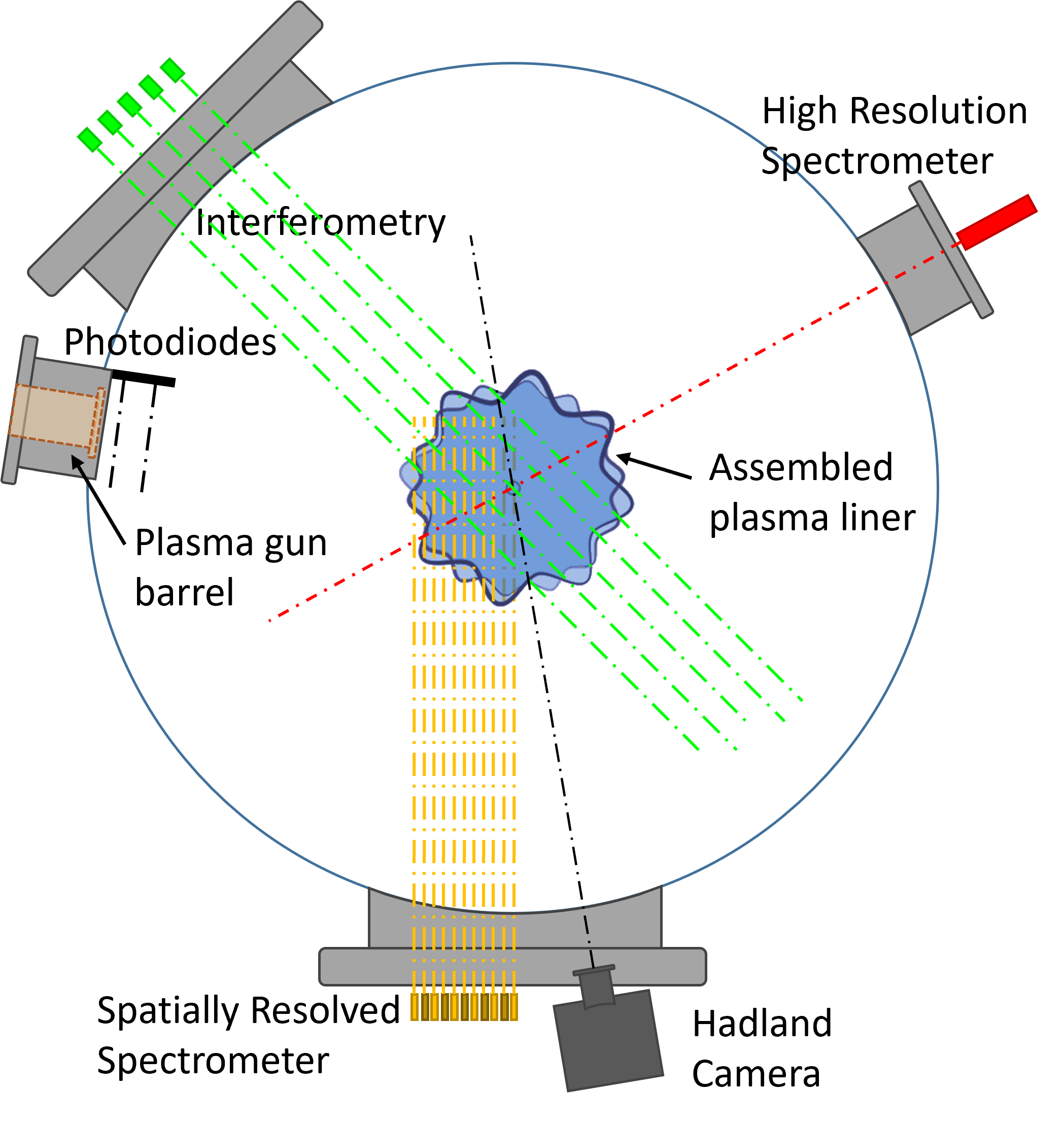}
\caption{\label{fig:AllDiagnosticsDiagram} Diagram of PLX and most of the central sight lines, indicated as dash-dotted lines, of the primary diagnostics. Spatial separations, angles, and placements are not necessarily to scale, and for ease of visualization, all diagnostics are shown here to be in the same 2D plane which is not the case on the actual 3D chamber.}
\end{figure}

A suite of diagnostics is used to interrogate the plasma liner properties throughout the merge and stagnation process. Time-resolved laser interferometry measurements of line-integrated plasma density are acquired over the duration of each shot, and a multi-frame fast camera images the plasma at several times throughout each shot. A collection of single-frame wide-view cameras are also used for imaging. Photodiode time-of-flight pairs are used measure individual jet velocities for each of the 36 plasma jets. Individual gun performance is also monitored using Rogowski coils that measure the main accelerating current of each gun (see Fig. \ref{fig:AllRogos_OnePlot_Shot2187}). An imaging spectrometer provides electron temperature and charge state information and high-resolution spectrometer is used to verify electron density and jet slowing at chamber center. Diagnostics lacking inherent time-resolution are instead triggered at varying times for different shots to build up an average picture of temporal evolution of the liner. Figure \ref{fig:AllDiagnosticsDiagram} indicates generally the layout and viewing arrangements of most of the diagnostics in the experimental campaign (the wide-view cameras are not indicated, as each has a view of the entire chamber interior).

\subsection{\label{sec:Diagnostics_Imaging}Imaging}

A multi-frame intensified Hadland UHSi24 camera in 12-frame mode and outfitted with a 24 mm focal length lens at f-number 11, provides a focused view of the chamber center where the ultimate liner stagnation occurs. A typical exposure time per frame of 25 ns is used with adjustable framerate set to 200 kHz. Additionally, a set of four single-frame PCO Pixelfly-USB cameras each outfitted with a 2.7 mm focal length fisheye lens at f-number 16 provide 1 $\mu$s exposure wide-angle images of the entire chamber and liner assembly and convergence process. These are set into the chamber via re-entrant ports described in detail in Ref[\onlinecite{myID_556}] so as to provide each with a view of the entire chamber interior. These Pixelfly cameras originally were intended to be used together (as a set of 6) to determine jet velocities of all 36 jets within a single shot \cite{myID_556}, but the photodiode method described in the next subsection proved to be a more reliable method for jet velocimetry.

\begin{figure}
\includegraphics[width=8.5cm]{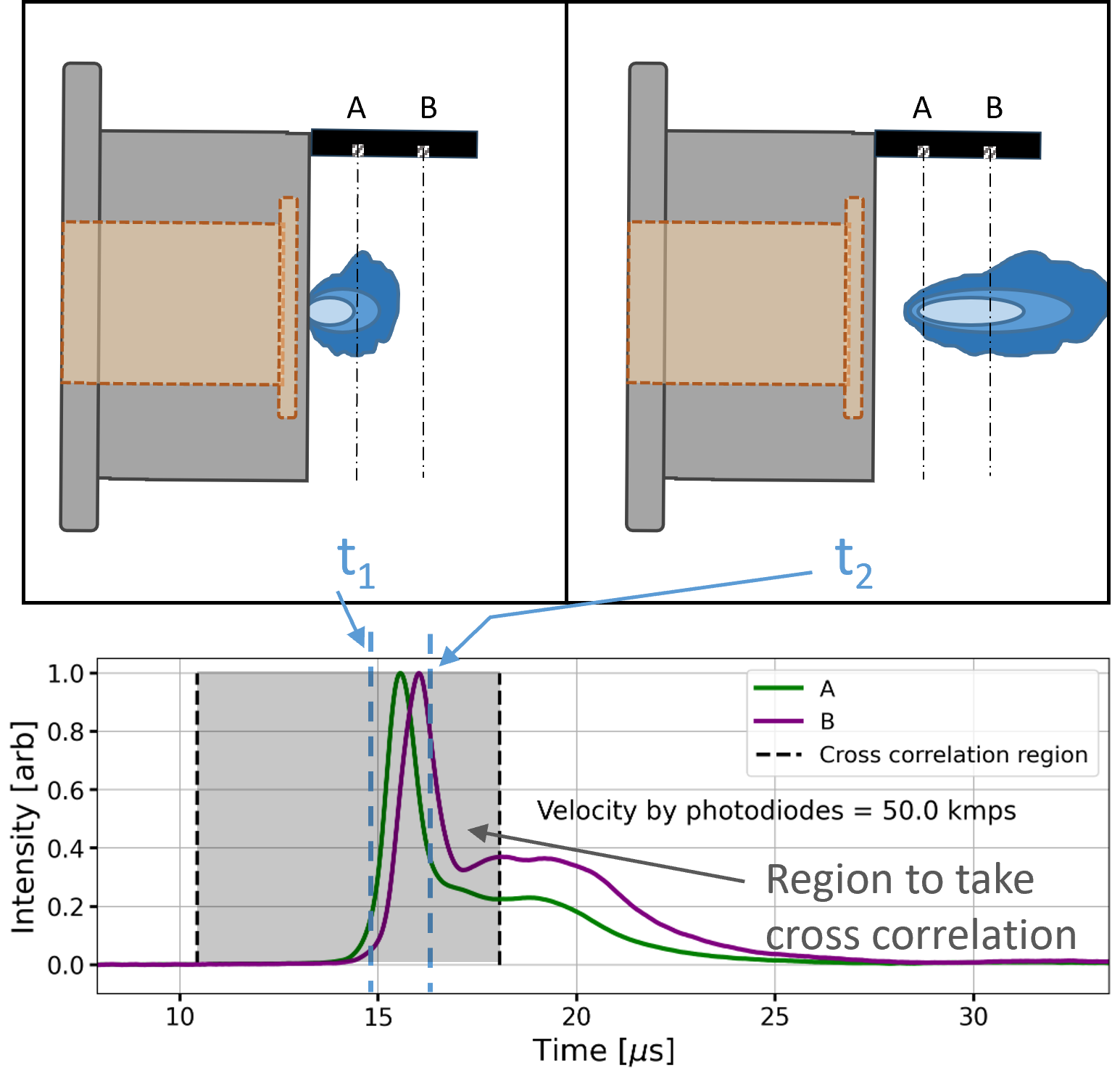}
\caption{\label{fig:PD_Diagram} Diagram illustrating photodiode time-of-flight pair operation.  The two diagrams represent the times $t_1$ and $t_2$, each indicated on a typical photodiode pair signal after smoothing. The velocity indicated is obtained via signal cross correlation within the highlighted region.}
\end{figure}

\subsection{\label{sec:Diagnostics_Photodiodes}Photodiode Pairs}
Each gun is equipped with two light-collecting fibers positioned along the jet trajectory and spaced by 2 cm which feed to two photodiodes having sufficiently fast risetime to track the jet (5 ns). The collection fibers are recessed into their holders by 6 mm to limit the light collection cone. The 2 cm separation distance was selected as a compromise between two competing considerations: there must be enough time separation between the signals that the available digitization rate of 40 MHz does not significantly contribute to uncertainty (i.e. there are sufficient samples separating the two signals), and the closer together the fibers are the signals are more likely to be closely correlated in shape which enables easier analysis. 

Figure \ref{fig:PD_Diagram} shows a diagram of the two photodiode views of a jet, at two times of jet emergence from the gun cavity. The corresponding locations on normalized sample signal readouts are also indicated, along with the region within which a cross correlation is applied. The bounds of the cross correlation region are selected to be a few microseconds before the signal rises, and a few microseconds after the signal peak. This provides the jet front velocity. The jets consist of a distribution of directed velocities with the trailing portion being slower than the jet front, resulting in an extended column of plasma. The velocity analysis focuses on the front region because this the portion which forms the spherical liner to compress the MIF target and is therefore most relevant. The photodiode analysis process is automated and repeated for each gun and enables velocities to be measured immediately following each shot and tuned as desired by adjusting the GV bank voltage - this we call the tuned velocity setting.

\begin{figure}
\includegraphics[width=8.5cm]{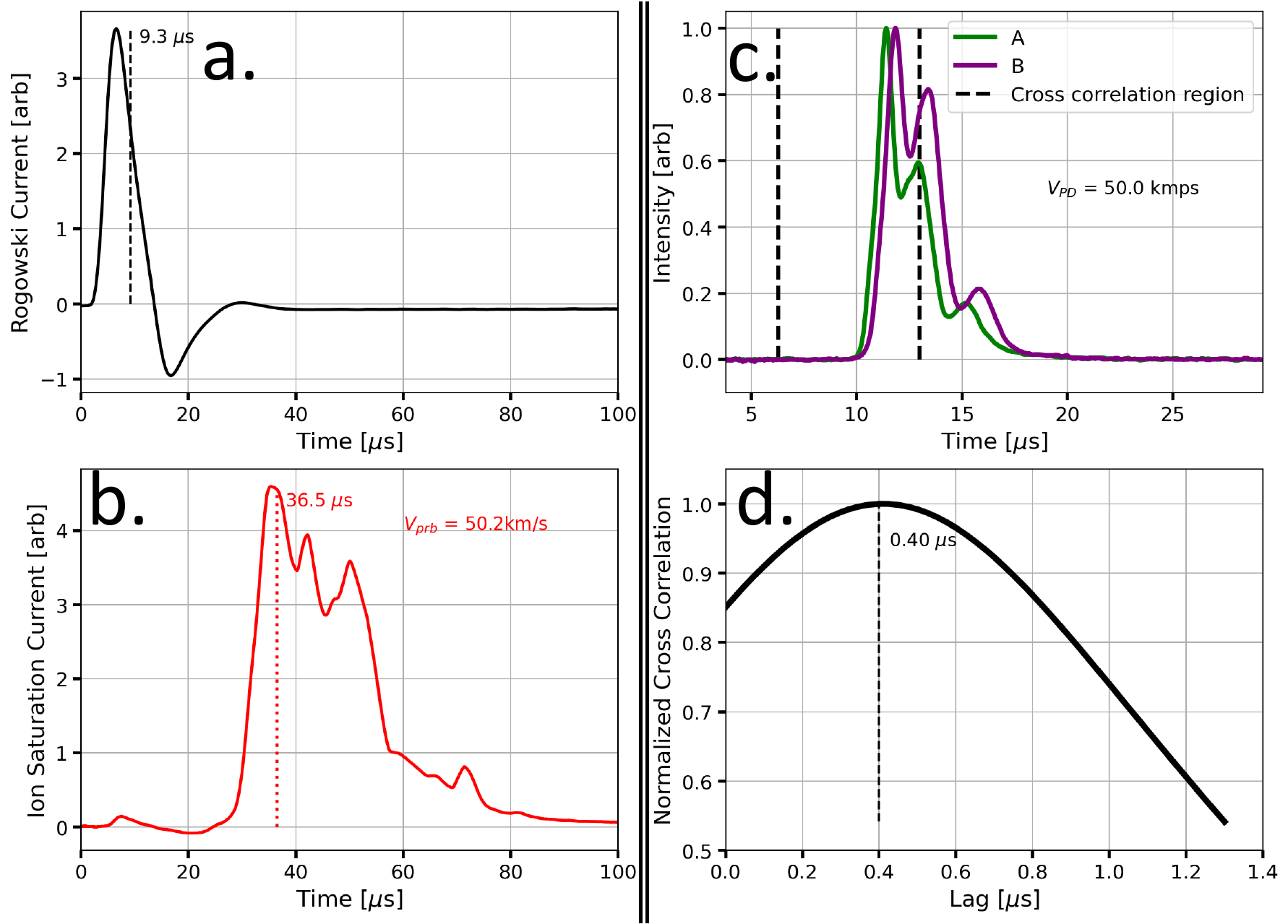}
\caption{\label{fig:PDCrossCorrelationResults} Comparison of jet velocity determined from a single-tip Langmuir probe and from the photodiode methodology described in Section \ref{sec:Diagnostics_Photodiodes}. The first column shows a) the Rogowski coil signal with jet ejection time and b) chamber-center probe signal with first-peak feature representing jet arrival time to probe. The second column indicates c) photodiode signals and d) cross correlation of those signals used to determine velocity.}
\end{figure}

This photodiode pair time-of-flight velocimetry methodology was cross-checked by taking a set of shots with a single gun at a time while positioning a single-tip Langmuir probe at chamber center, and obtaining a jet velocity via the plasma time-of-flight to the probe ($V_{prb}$) and comparing to the velocity inferred via the photodiode signals ($V_{PD}$). A typical comparison is shown in Fig. \ref{fig:PDCrossCorrelationResults}. In this, we define the jet arrival time to the probe as the location corresponding to 20\% of the full integral of the probe signal. The previously discussed jet ejection time is subtracted from this arrival time, and in conjunction with the travel distance being the PLX radius of 136.5 cm, provides $V_{prb}$. This generally matches well with the first peak-like feature in the signal, as seen in subfigure b. 

It was found that even with a careful analysis methodology, the photodiode velocimetry incurred additional shot-to-shot variation in inferred velocity compared to the probe time-of-flight technique. This is believed to be due to early time spatial structure and inhomogeneity in the plasma jet. A tune setting was selected for final use if the mean velocity measured on the photodiodes was 50 $\pm$ 10 km/s over about 5 shots.

\subsection{\label{sec:Diagnostics_HiResSpect}High Resolution Spectroscopy}
Single-emission-line shapes are acquired using a 2 m double-pass (for a 4 m focal length) high-resolution monochromator (McPherson 2062DP) with 2400 mm$^{-1}$ grating and 50 $\mu$m slit width, with the resulting spectrum captured with a 1 $\mu$s exposure on a PCO Pixelfly camera. The emission studied in this work is the Ar II 480.60 nm line due to its relatively high intensity among the other visible band Ar II emission lines, in addition to a lack of local lines from higher ionization states or impurities. The line of sight (LOS) of this spectrometer is directly through PLX chamber center.

For this plasma emission line profile, there are three primary plasma factors that we find to impact the observed shape. Bulk motion of the plasma throughout the LOS causes Doppler shift of the observed emission wavelength, and a distribution of bulk plasma motion results in a corresponding distribution of shifts. Doppler broadening associated with the plasma temperature of this distribution is also present. Finally, Stark broadening associated with the electron density in the vicinity of emitters also impact the observed shape. 

At the stagnation densities of $\sim10^{17}$ cm\textsuperscript{-3}, the time required for equilibration of electron and ion temperatures is far shorter than the $\sim \mu$s timescales over which the plasma evolves. We assess this using the Coulomb collision expression for temperature equilibration given in Eq. \ref{eq:TemperatureEquilibration} in which species $\alpha$ corresponds to ions and $\beta$ to electrons, and $\bar{\nu}_\epsilon^{\alpha\backslash\beta}$ is the energy-exchange frequency in units of s\textsuperscript{-1} between electrons and ions given in Eq. \ref{eq:EnExFrequency}:
 
\begin{equation}\label{eq:TemperatureEquilibration}
\frac{dT_\alpha}{dt} = \bar{\nu}_\epsilon^{\alpha\backslash\beta} (T_\beta - T_\alpha)
\end{equation}

\begin{equation}\label{eq:EnExFrequency}
\bar{\nu}_\epsilon^{\alpha\backslash\beta} = \num{1.8e-19} \frac{(m_\alpha m_\beta)^{1/2}Z_\alpha^2 Z_\beta^2 n_\beta \lambda_{\alpha\beta}}{(m_\alpha T_\beta + m_\beta T_\alpha)^{3/2}} 
\end{equation}

\noindent With typical expected conditions of $T_e = 2.5$ eV, $n_e = \num{3e17}$ cm\textsuperscript{-3}, and supposing ion temperature $T_i$ is moderately elevated at $10$ eV, the resulting time required for temperature equilibration between electrons and ions is on the order of nanoseconds. Thus, $T_i \approx T_e = T$ and are also clamped to values around 2.5 eV - as indicated by broadband emission spectroscopy results in Section \ref{sec:Results_Spectroscopy}. Thus ultimately, Doppler broadening is found to be negligible in comparison to the other two broadening mechanisms, meaning that the high resolution lineshapes are more useful for density and residual velocity inference than for temperature determination.  

\subsection{\label{sec:Diagnostics_Interferometry}Multi-Chord Interferometry}

Line-integrated electron density is obtained in this work via a multi-chord heterodyne laser interferometry system whose arrangement is laid out in Ref[\onlinecite{myID_517}]. The laser emission is produced using a 320 mW, 561 nm solid state laser, is split into multiple lower-power beams and injected into the chamber through five single-mode fiber optic cables with each probe beam about 0.3 cm in diameter. The scene chords of the interferometer are aligned parallel to one another, with a central chord passing through the center of the chamber, and the others in pairs, offset from the central by 12.75 cm and 25.5 cm respectively. Thus the chords measure the coarse radial density profile as well as provide a check against the spherical symmetry of the liner (if a perfectly spherical plasma liner is formed, the signals on the chord pairs should perfectly match). After passing through the plasma, the probe beams are compared with a reference beam of the interferometer in the Mach-Zender configuration, and the line-integrated electron density is computed based on the relative phase shift \cite{myID_555,myID_517}.

\subsection{\label{sec:Diagnostics_SpatiallyResEmSpec}Spatially-Resolved Emission Spectroscopy}
Spatially-resolved visible plasma emission spectra provides insight into LOS-integrated behavior of electron temperature via relative line emission intensities, and relative plasma density via observed Bremsstrahlung emission intensity. In the Multi-Chord Spectrometer (MCS), data is collected along 11 fiber-coupled lines of sight through 2.5 cm diameter collimating lens of 6 cm focal length mounted outside the chamber, with waist size of 1 cm. This light is focused on the entrance slit of a Chromex 500is imaging spectrometer with 0.5 m focal length, a 1200 grooves/mm grating and 50 $\mu$m slit for an effective resolution of 0.08 nm over the wavelength range used, 453.3 nm to 482.2 nm. A relative intensity calibration was applied based on a calibrated tungsten Thorlabs SLS201L lamp shining through the same set of windows and optics as the PLX plasma light traverses. An additional calibration measure was taken to normalize the total amount of light collected through each fiber; to do so, each fiber was illuminated individually with the lamp and imaged with constant camera settings, allowing for comparison of counts vs. wavelength for all fibers.

\begin{figure}
\includegraphics[width=8.5cm]{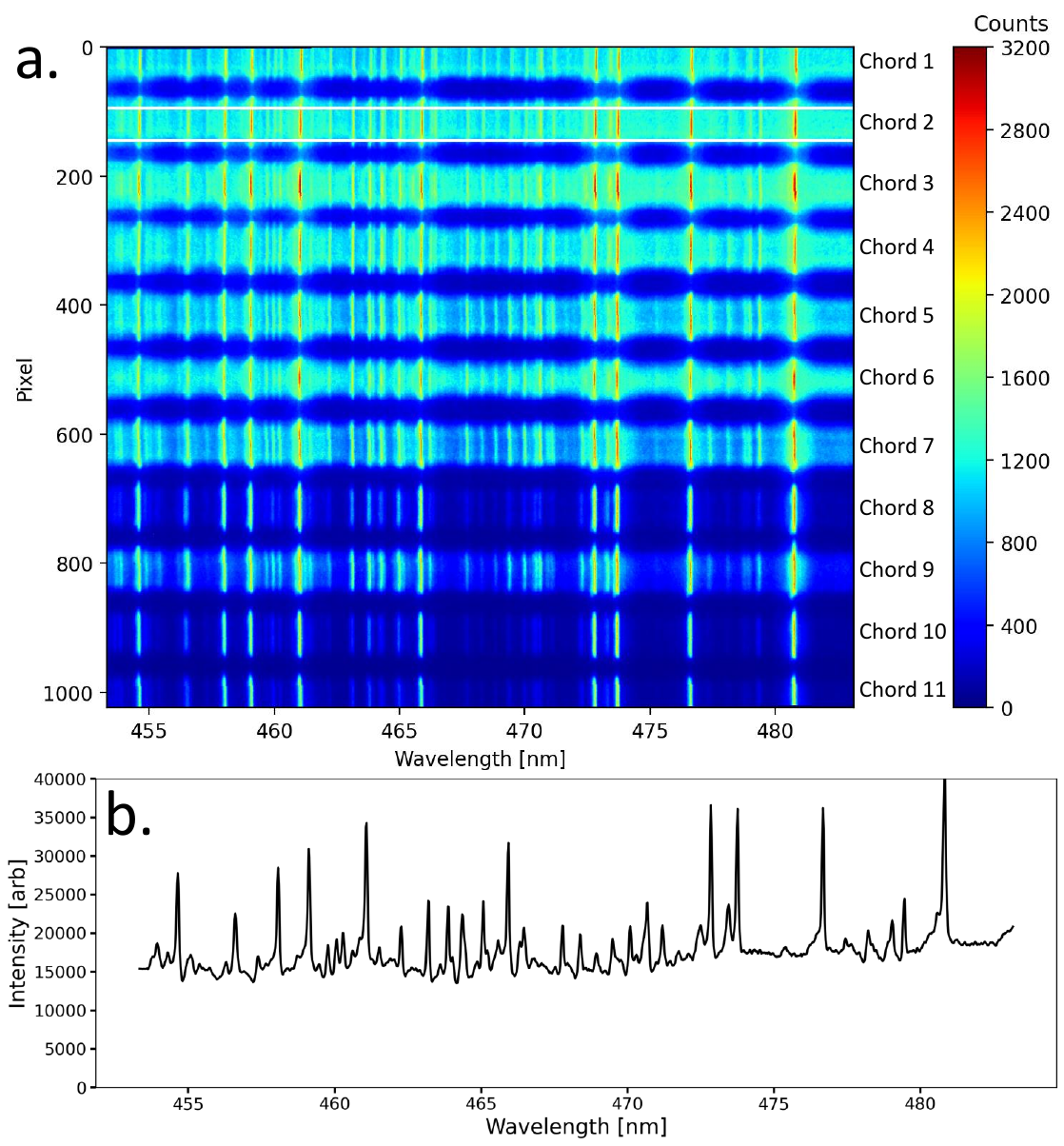}
\caption{\label{fig:MCS_ConsolidatedResults_abParts} a) MCS spectral image obtained with 1 $\mu$s exposure taken 45.4 $\mu$s after jet ejection time, around stagnation. Chord 2 (second-closest to chamber center) region is highlighted in white. b) Spectrum of chord 2 region, summed along vertical pixels within the highlighted region.}
\end{figure}

A sample spectral image is shown in Fig. \ref{fig:MCS_ConsolidatedResults_abParts}, as well as a spectrum corresponding to the highlighted region of fiber 2 after vertical summing of region signal counts and applying the intensity calibration. This process is repeated for respective fiber regions to obtain the 11 chordal spectra per image. The presence or lack of emissions from the Ar ionization stages constrains the temperature regime to electron temperatures of $\sim$few eV. More precisely, the best-fit line-integrated electron temperature is inferred by comparing the observed emission spectrum to a set synthetic spectra of varying temperature generated by the collisional-radiative spectral analysis model PrismSPECT. PrismSPECT generates an emission spectrum based on excited state densities that it calculates via a collisional-radiative model in a non-local thermodynamic equilibrium regime. The atomic level scheme utilized in the model includes excited states up to Ar IV, after which only ground states of ionic stages are considered (and are all of negligible populations).

The measured-to-synthetic spectra comparison is made after having subtracted off the continuum background from the MCS spectra. The PLX liner sweeps up residual air in the chamber as it travels inwards, so the observed spectra contain N and O lines in addition to the Ar lines, though no other contaminants are observed. The different chords of the MCS observe different amounts of these N and O lines relative to the Ar lines, so modeled spectra are generated for different elemental compositions as well. After the entire set of composite synthetic spectra at varying $T_e$ and compositions are generated, a simple least-squares method is performed to select that which matches closest to the observed MCS spectrum for a given chord, and we take that $T_e$ as the LOS-integrated representative temperature for that chord.

Another characteristic that the MCS provides is LOS-averaged plasma density relative to that of the central chord (this doesn't involve comparison to PrismSPECT). This is a relative measurement due to lack of absolute intensity calibration with the spectrometers. Not considering all factors which do not depend strongly on density, the free-free (Bremsstrahlung) emissivity scales as:

\begin{equation}\label{eq:freefree}
\varepsilon_{ff} \propto \sum_i n_e Z_i^2 n_i
\end{equation}

\noindent and the free-bound (radiative recombination) emissivity scales as:

\begin{equation}\label{eq:freebound}
\varepsilon_{fb} \propto \sum_i  \frac{n_e Z_i^2 n_i}{U_i}
\end{equation}

\noindent in which $U_i$ is its partition function \cite{myID_561,myID_562}.

Our plasma is primarily comprised of Ar\textsubscript{+} and Ar\textsubscript{2+}, so electron density $n_e \approx n_{+} + 2n_{2+}$. Given this, the combination of free-free emission and free-bound radiation scales as $n_e^2$, so we need only have a reasonable knowledge of $n_e$ and mean charge state $\bar{Z} \approx \frac{n_e}{n_{+}+n_{2+}}$ to have the necessary scaling relation between continuum intensity to spatially-resolved relative plasma density. These characteristics are quantified in Section \ref{sec:Results_Spectroscopy}. A notable feature discussed in that section is the presence of N and O constituents in the plasma which add their own contributions to $n_e$ and thus complicate the scaling behavior in $\epsilon_{ff}$ and $\epsilon_{fb}$ - however, we will still retain the relation $(\epsilon_{ff} + \epsilon_{fb}) \propto n_e^2$ as an approximation.

\section{\label{sec:Results}Results\protect\\}

\begin{figure*}
\includegraphics[width=18.2cm]{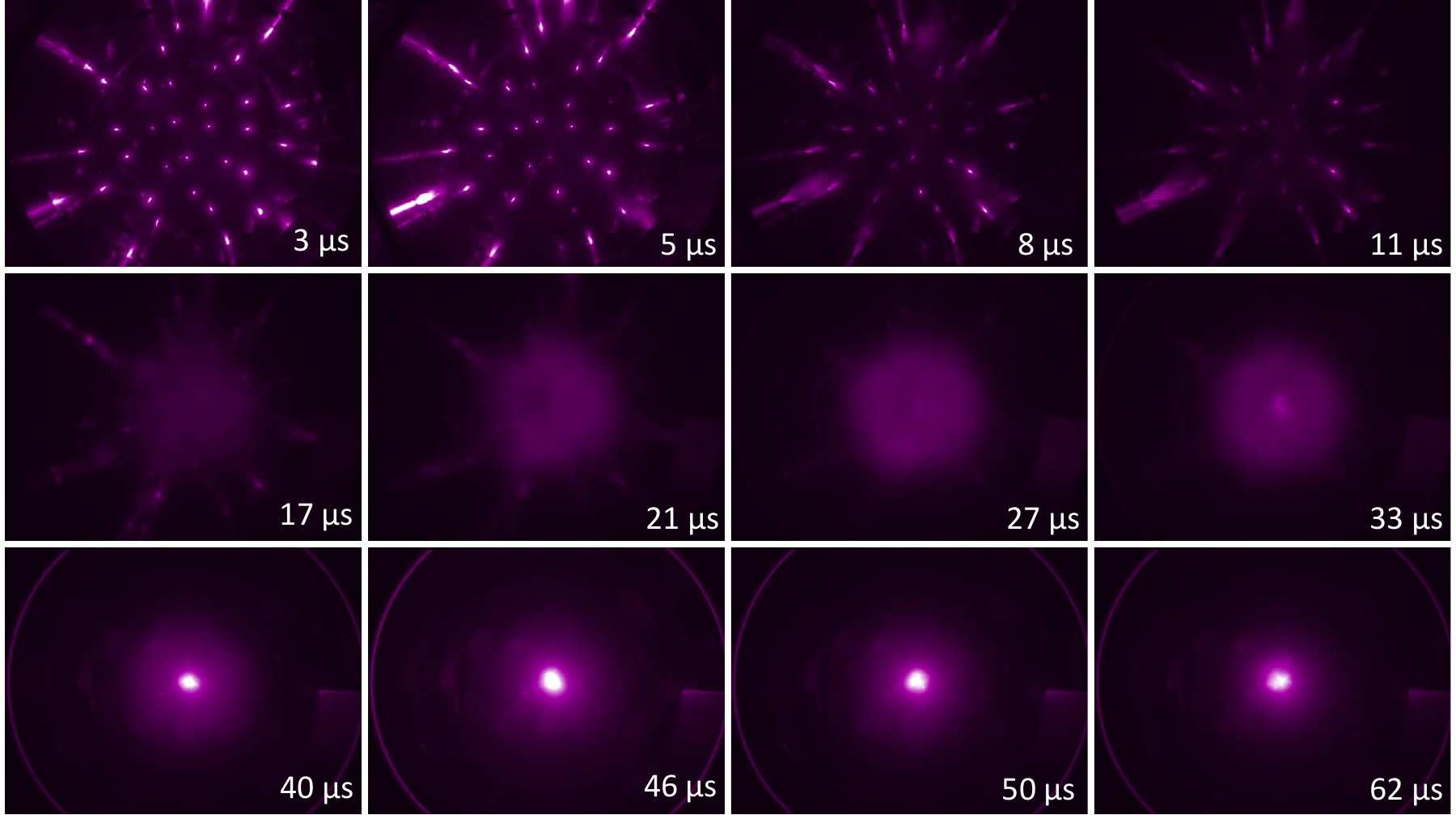}
\caption{\label{fig:LinerMergeProgression_font2}False color images from a series of different shots at different times throughout the jet ejection and merge process from one of the fisheye cameras, with counts shown on a logarithmic scale.}
\end{figure*}

\subsection{\label{sec:Results_Imaging}Imaging}
\begin{figure}
\includegraphics[width=8.5cm]{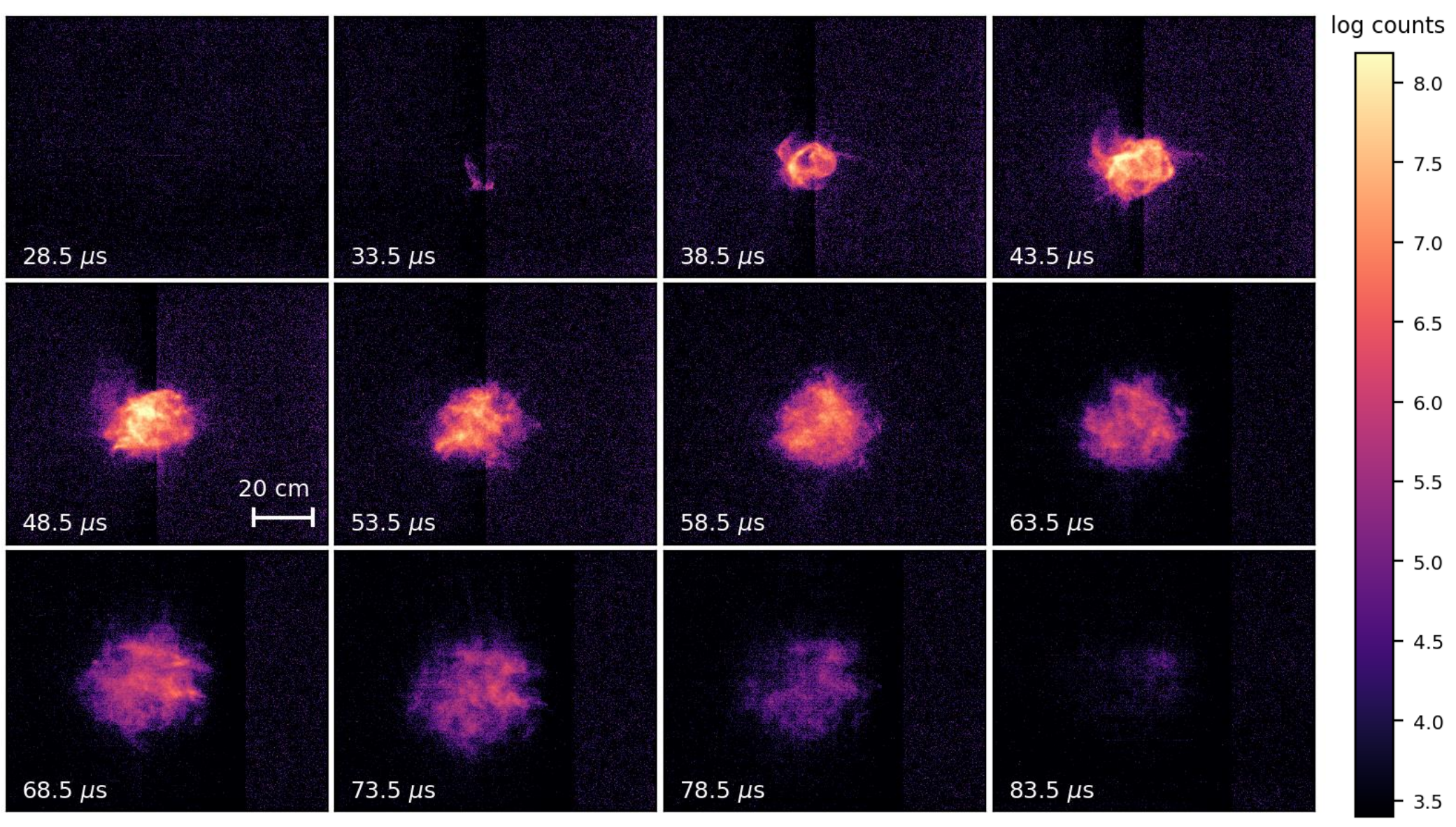}
\caption{\label{fig:HadlandImages2187_cmapMagma} Single shot image set from the multi-frame Hadland ICCD camera. Image times after ejection are included as well as the spatial scale. Exposures are 25 ns, and image counts are shown on a heatmap with a logarithmic scale. This shot is the same as that for which Rogowski coil signals were shown in Fig. \ref{fig:AllRogos_OnePlot_Shot2187}.}
\end{figure}

A series of images from the Pixelfly fisheye cameras is shown Fig. \ref{fig:LinerMergeProgression_font2}, and illustrates the general behavior from jet ejection to initial merge and finally to central stagnation. The liner assembly begins between 11 and 17 $\mu$s. Stagnation and the brightest moment occurs around 46 $\mu$s. It should be noted that because each image corresponds to a different shot, jet velocities may deviate slightly between each image as a result of shot-to-shot variation (see Fig. \ref{fig:ShotToShotVariation}). In contrast, the set of images shown in Fig. \ref{fig:HadlandImages2187_cmapMagma} are from a single shot using the multi-frame Hadland camera. Both the Pixelfly images and Hadland images indicate a uniform liner is indeed formed by the 36 discrete jets with shockless morphology en route to chamber center, as opposed to earlier observations using a smaller subset of guns \cite{myID_549,myID_570} operating at different conditions. Hadland multi-frame imaging of the process for a single shot also shows a liner stagnation which maintains good sphericity, deviating slightly in some frames towards a more oblate shape. This may be related to the 9 TL guns on the bottom of the chamber which necessarily fire with slightly different settings than the other guns as discussed in Section \ref{sec:Experiment}. The dissipation time that the overall emission strength recedes back to background levels is around 40 $\mu$s after stagnation.

\begin{figure}
\includegraphics[width=8.5cm]{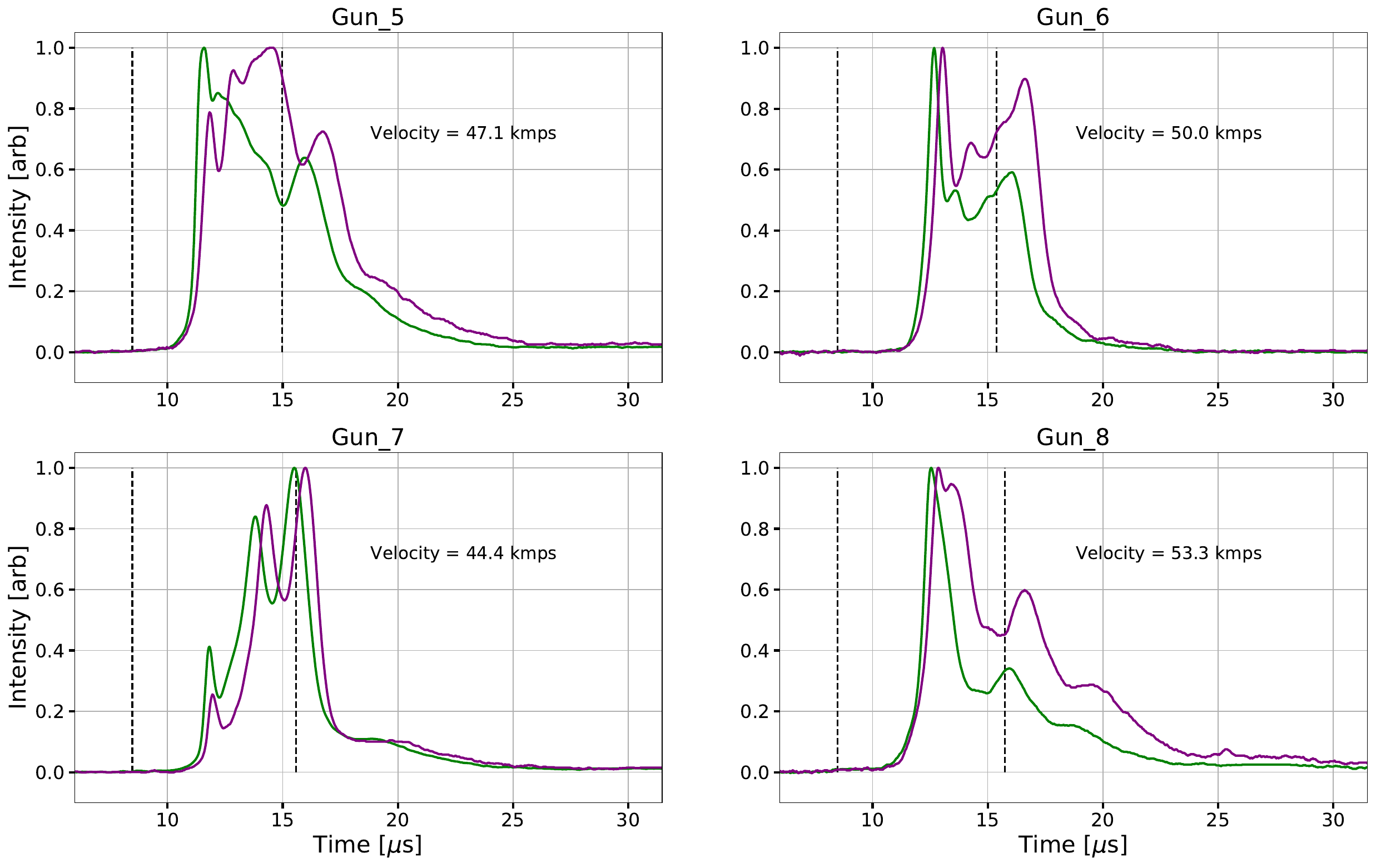}
\caption{\label{fig:PD_Shot2187_Quad2} Photodiode signals for several guns and the resulting velocity from cross correlation within the marked region. Despite the diversity of signal shapes, velocities are relatively consistent.}
\end{figure}

\subsection{\label{sec:Results_VelocityDetermination}Jet Velocities and Liner Merging}
Photodiode signals for several guns are shown in Fig. \ref{fig:PD_Shot2187_Quad2} in which one observes a diversity of photodiode signal shapes observed. One potential source of the form variation is that the jets are expelled from the gun with more turbulent structure than anticipated. This variation in shapes practically means that certain jets have their velocities determined more reliably than others.

\begin{figure}
\includegraphics[width=8.5cm]{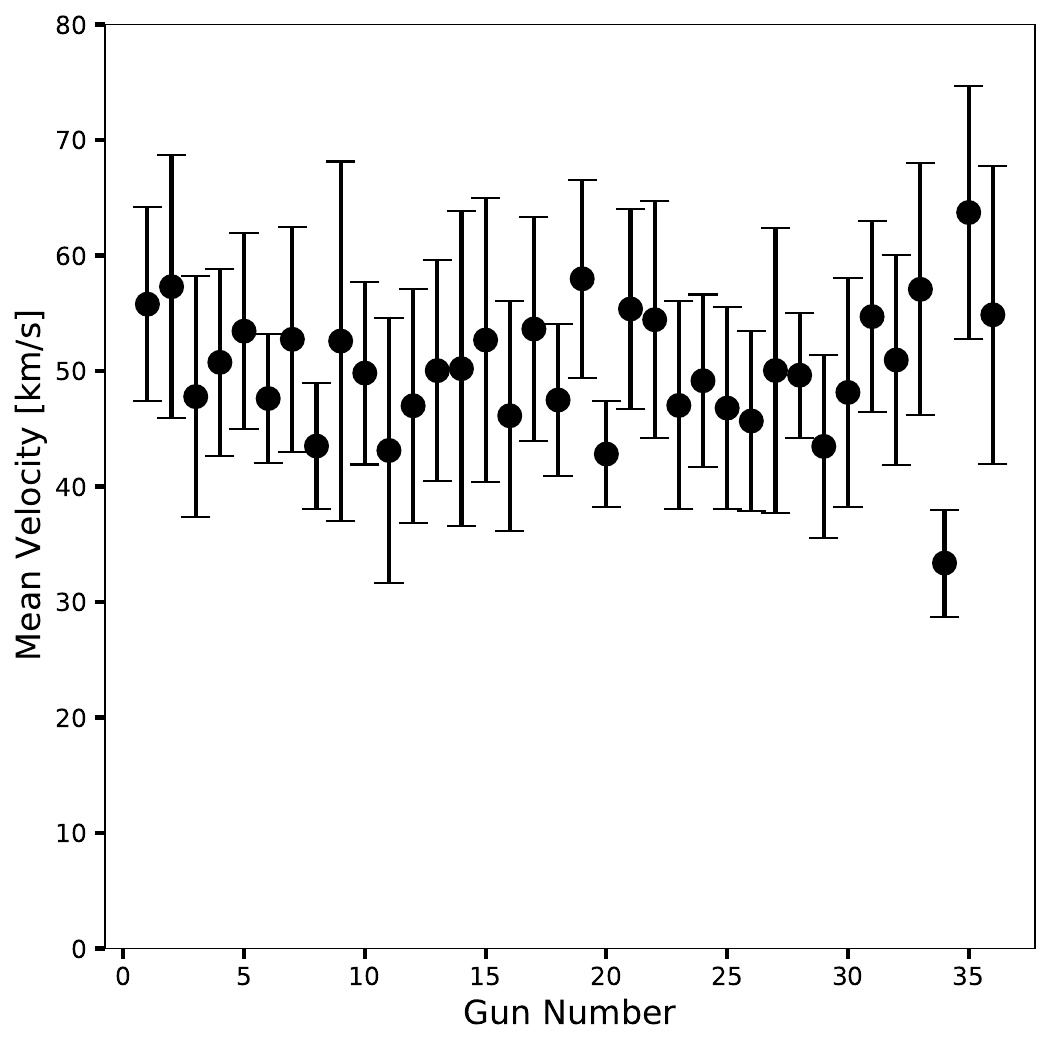}
\caption{\label{fig:GunVelocities_AvgOverShots_Velocity_BaseSig_m2} Mean velocities and standard deviations (error bars) of all 36 guns, averaged over a set of 30 shots over which tune values are unchanged and selected to nominally obtain 50 km/s jet front speeds for all jets.}
\end{figure}

\begin{figure}
\includegraphics[width=8.5cm]{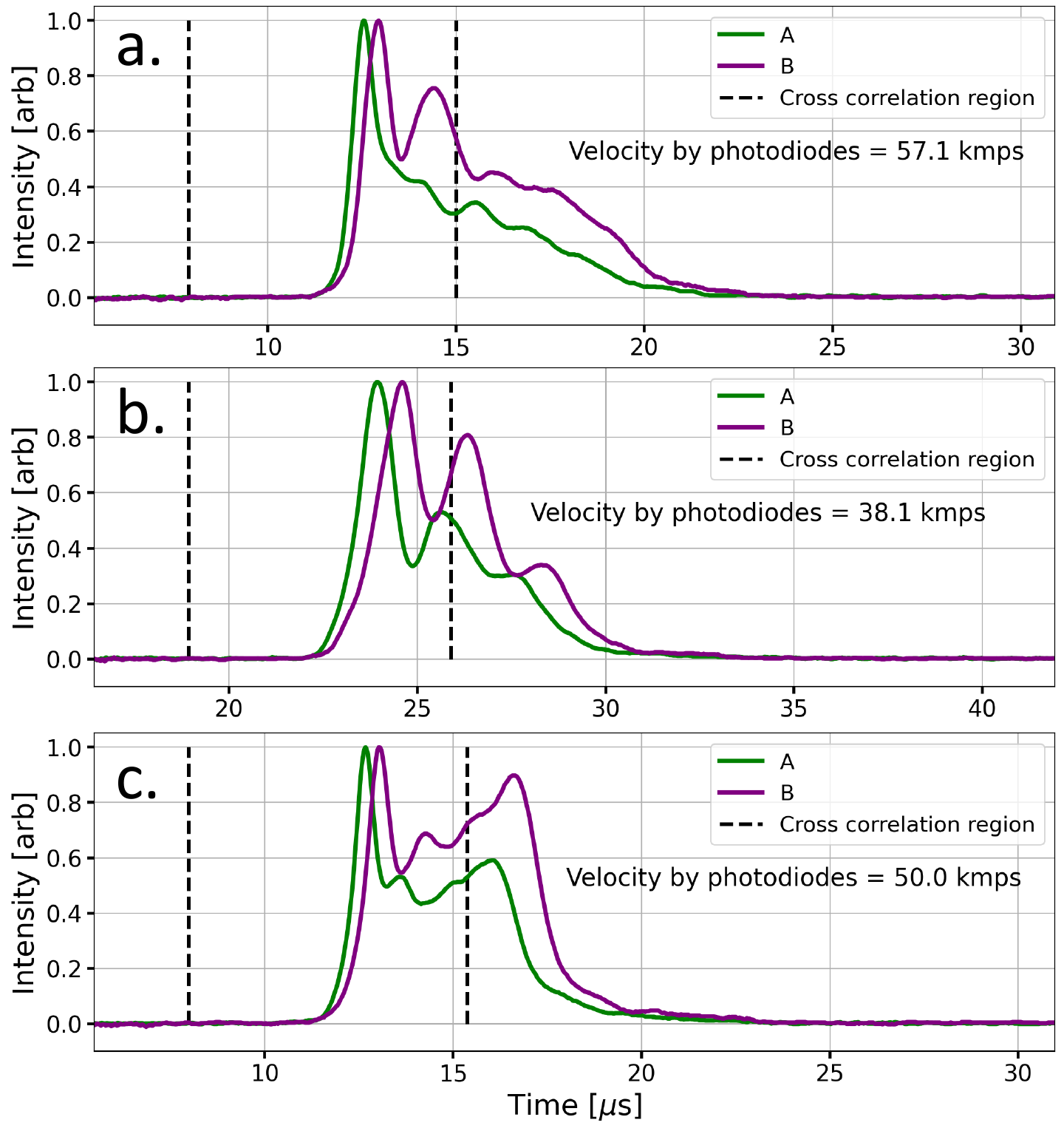}
\caption{\label{fig:ShotToShotVariation} Photodiode signals from a single gun for three different shots, with three different calculated velocities, indicative of shot-to-shot variation.}
\end{figure}

Velocities averaged over a set of 30 shots are shown in Fig. \ref{fig:GunVelocities_AvgOverShots_Velocity_BaseSig_m2}. Uncertainties are more so a result of shot-to-shot variation rather than in the analysis methodology. This variation is visualized in Fig. \ref{fig:ShotToShotVariation}, in which the three shots shown give rather different velocity values, which can be qualitatively confirmed by noting the separation of the two signals in e.g. subplot a. versus those in b. Although shot-to-shot variation is evident in the magnitude of the error bars, mean values over the shot series generally fall within $\pm$ 10 km/s of the targeted 50 km/s with few outliers - the tuning of gun 34 for example was ineffective as the reduction in GV voltage necessary to elevate its velocity also prevented its firing at all. The velocity determination method discussed in Section \ref{sec:Diagnostics_Photodiodes} returns clearly erroneous velocity values in about 10\% of cases (by-hand calculation based on signal separation in these cases sharply disagrees with those calculated via the cross correlation method). To discard these as outliers, for a given jet on a given shot only calculated front velocities between 20 and 85 km/s are accepted. With present gun settings it is unlikely jets are fired at velocities truly outside these bounds.

\subsection{\label{sec:Results_Spectroscopy}Spatially-Resolved Emission Spectroscopy}

\begin{table}
\caption{\label{tab:PSpectParams}PrismSPECT parameters corresponding to those in Fig. \ref{fig:Decomposition_BestMatch_Shot2170_Fiber1}.  }
\begin{ruledtabular}
\begin{tabular}{l l}
Model Property & Value (Ar / N / O) \\
\hline
Overall ion density & $\num{1.0e17}$ cm\textsuperscript{-3} \\
Plasma thickness & 15 cm \\
Matched electron temperature & 2.5 eV\\
Matched composition  & 0.6 / 0.18 / 0.22 \\
Matched mean ion charge & 1.8 / 1.4 / 1.2
\end{tabular}
\end{ruledtabular}
\end{table}

\begin{figure}
\includegraphics[width=8.5cm]{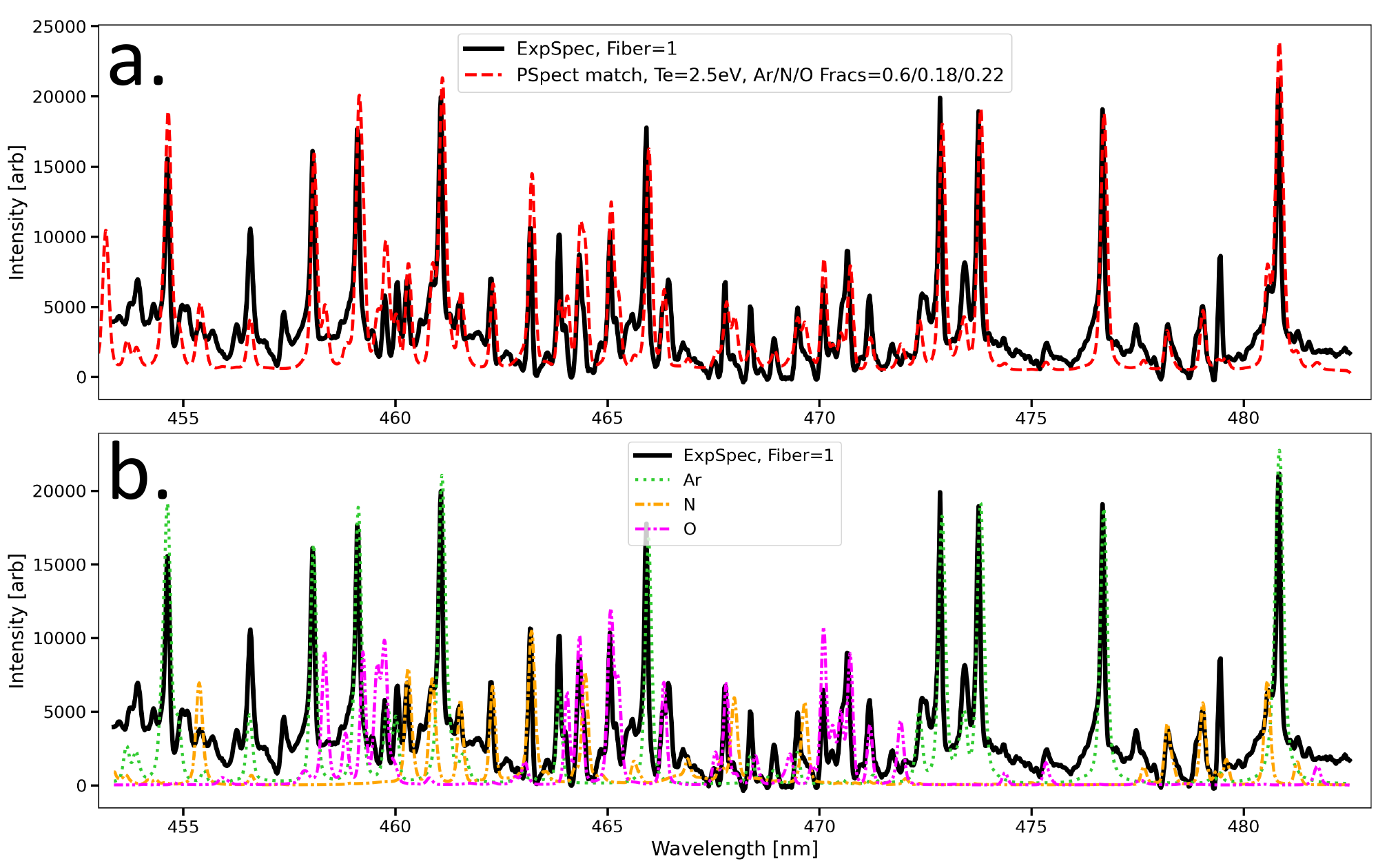}
\caption{\label{fig:Decomposition_BestMatch_Shot2170_Fiber1} a) Sample experimental spectrum with the best matched of a series of synthetic spectra of mixed species plasma. b) Same experimental spectrum and independent synthetic spectra of Ar, N, and O constituents illustrates which observed lines belong to which species. These are from the same spectral image in Fig. \ref{fig:MCS_ConsolidatedResults_abParts}, after having subtracted the background.}
\end{figure}

While the combination of shot-to-shot velocity variation and jet-to-jet variation within a single shot constrains attainable liner uniformity, it remains sufficiently minimal to provide some interesting results throughout the merge process by examining a collection of shots. Spectra observed with the MCS spatially-resolved spectroscopy system indicates a marked rise in presence of air lines closer to the central chord (chord 1). A sample spectrum from the central chord of the MCS is shown in Fig. \ref{fig:Decomposition_BestMatch_Shot2170_Fiber1}, along with synthetic PrismSPECT spectra generated using the characteristics denoted in Table \ref{tab:PSpectParams}. The best matching synthetic spectrum of a set with varying properties is the red curve in subfigure b., corresponding to an electron temperature of 2.5 eV, an Ar fraction of 0.6, and the remaining composition split nearly equally between N and O species. That N and O are present in roughly equal quantities instead of a fraction similar to air suggests that some oxygen is desorbed from plasma-facing surfaces in the chamber during a shot, elevating it more than the ordinary air fraction. The fraction of N and O increases for the LOS closer to chamber center. Mean charge state $\bar{Z}$ is another parameter determined via this PrismSPECT comparison, with argon $\bar{Z} \approx 1.8 \pm 0.14$ for the above conditions.

\begin{figure}
\includegraphics[width=8.5cm]{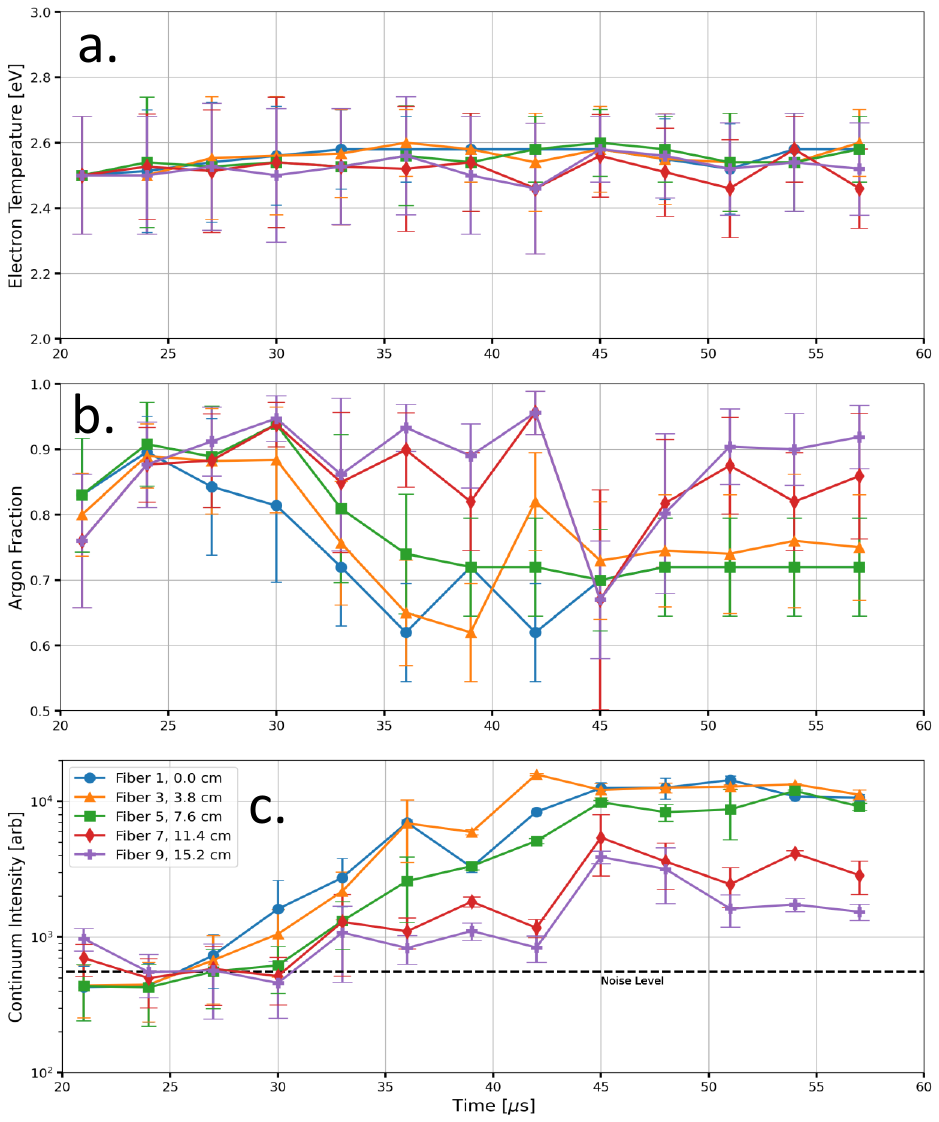}
\caption{\label{fig:MCS_ConsolidatedResults_cdeParts2} Results as obtained by the MCS, averaged over a set of 30 shots and binned into 3 $\mu$s intervals. a) Electron temperature determinations via PrismSPECT comparison across several chords (separation from central chord in cm is indicated). b) Fraction of Ar determined via PrismSPECT comparison. c) Continuum emission levels, correlated with plasma density.}
\end{figure}

Performing the comparison to PrismSPECT for all fibers at times throughout liner formation and implosion provides a number of interesting insights, summarized in Fig. \ref{fig:MCS_ConsolidatedResults_cdeParts2}. There is a center-wards (lower fiber number) decrease in Ar amount, observed in subfigure b. This decrease is not attributable to changes in electron temperature $T_e$ which remains at about 2.5 $\pm$ 0.18 eV over space and time as indicated in subfigure a., and is instead evidence of the residual chamber air being swept up by the liner front. The unchanging nature of $T_e$ was also observed in an earlier PLX study involving merging of only a few guns \cite{myID_549}. Another striking feature of the spectra is evolution of continuum emission intensity from increasing Bremsstrahlung and bound-free emissions illustrated in subfigure c. This provides relative plasma density behavior spatially and temporally, as the continuum intensity scales as $n_e^2$, as discussed in Section \ref{sec:Diagnostics_SpatiallyResEmSpec}.

\subsection{\label{sec:Results_HiResSpectroscopy}High Resolution Spectroscopy}
The high resolution spectrometer described in Section \ref{sec:Diagnostics_HiResSpect} captures a finely resolved profile of the Ar II 480.60 nm emission line. This profile shape originates through a convolution of several mechanisms - the easiest of these to analyze is the instrumental profile whose typically Gaussian shape is the profile constituent originating from optical components, not the plasma itself. Stark broadening manifests as a Lorentzian profile in response to the electric microfield from electrons and ions in the vicinity of emitting ions, and therefore scales linearly with the plasma density. Associated Stark widths for different densities for this line are obtained in Ref. \onlinecite{myID_560}. The last profile contribution is a composite effect from each jet conferring a Doppler shifted profile with the previously mentioned broadening freatures due to the projection of the jet's radial velocity onto the spectral LOS.

To analyze the share of these attributes in the observed emission profile, a technique was developed that models the incoming jets as cylinders of uniform diameter and singular speed rather than a distribution of speeds. These cylindrical jets intersect the instrument's LOS which itself is approximated as a narrower cylinder at an angle corresponding to the gun port location relative to the collection optics port. As a result, jets originating close to the collection optics port and close to the polar opposite location will contribute a greater proportion of light to the overall profile because they have a greater intersection volume, and will also be Doppler shifted the most due to the shallow angle of jet with respect to optics LOS. Using this framework, resultant composite profiles can be reconstructed for a set of values of plasma bulk velocities and densities and seeing which best match the observed data (bearing in mind that the simplicity of the analysis model will contribute some uncertainty).

\begin{figure}
\includegraphics[width=8.5cm]{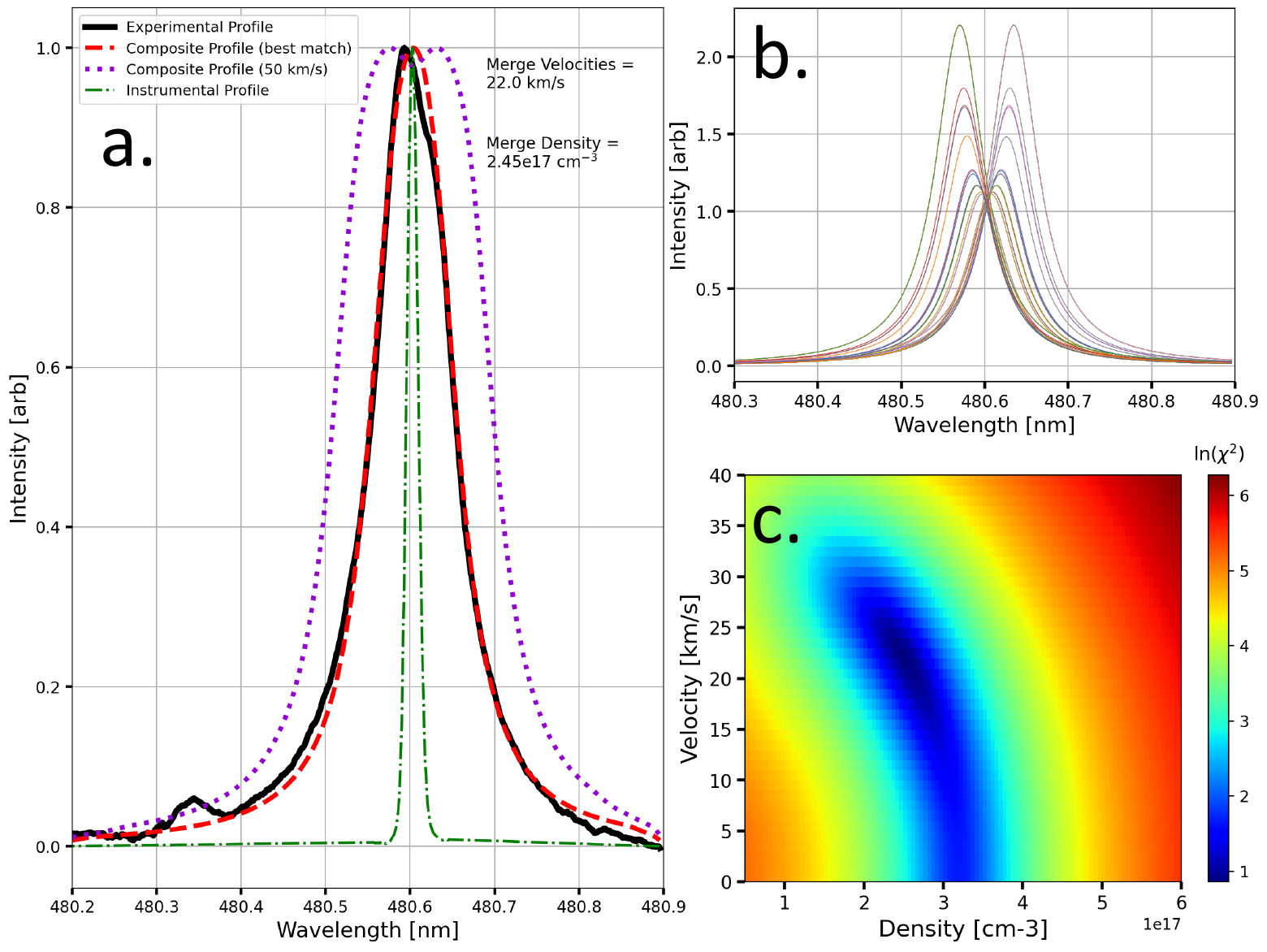}
\caption{\label{fig:HiRes_Shot2170Analysis} High-Resolution profile analysis for a shot at 45.4 $\mu$s after ejection with 1 $\mu$s exposure. a) Observed emission profile of Ar II 480.6 nm line, along with the best matching reconstructed profile (jet merge speeds and density are indicated), a poor match assuming 50 km/s speeds, and the instrumental profile. b) Individual profiles for the best match and amplitude scalings from each jet based on angle with respect to collection optics. c) Contour map of goodness-of-fit factors for a set of modeled merge densities and velocities - the lower value indicates better matching.}
\end{figure}

Carrying out this analysis on a shot very near stagnation time, the results are given in Fig. \ref{fig:HiRes_Shot2170Analysis}. Subfigure a. shows the experimentally observed profile, the reconstructed profile with the indicated parameters, and the instrumental profile obtained by looking at the H$\beta$ emission line using a low pressure discharge hydrogen lamp. Subfigure b. indicates the individual component profiles corresponding to each jet - note the varying amplitude and Doppler shifts depending on the gun's placement relative to the collection optics. Subfigure c. shows the quality of match between the reconstructed profile and the experimental one (lower values indicate better matching) as a function of model parameters of jet speeds and plasma density. A particularly interesting result is that to capture the observed profile, we must reconstruct using velocities about 20 km/s or less, much less than our ejected jet velocities. Indeed, using a velocity closer to the ejection velocities of 50 km/s produces not only a profile which tends to be too broad, but also is double-peaked with a valley in between (see purple curve in Fig. \ref{fig:HiRes_Shot2170Analysis}), features not observed in experimental data. 

\subsection{\label{sec:Results_Interferometry}Interferometry}

\begin{figure}
\includegraphics[width=8.5cm]{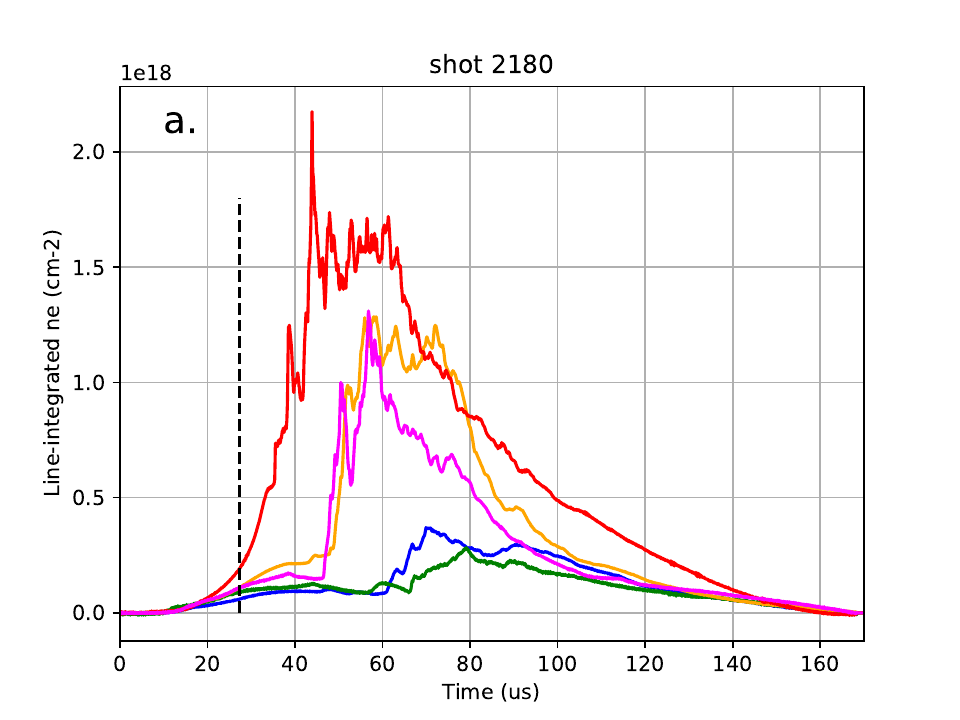}\\
\includegraphics[width=8.5cm]{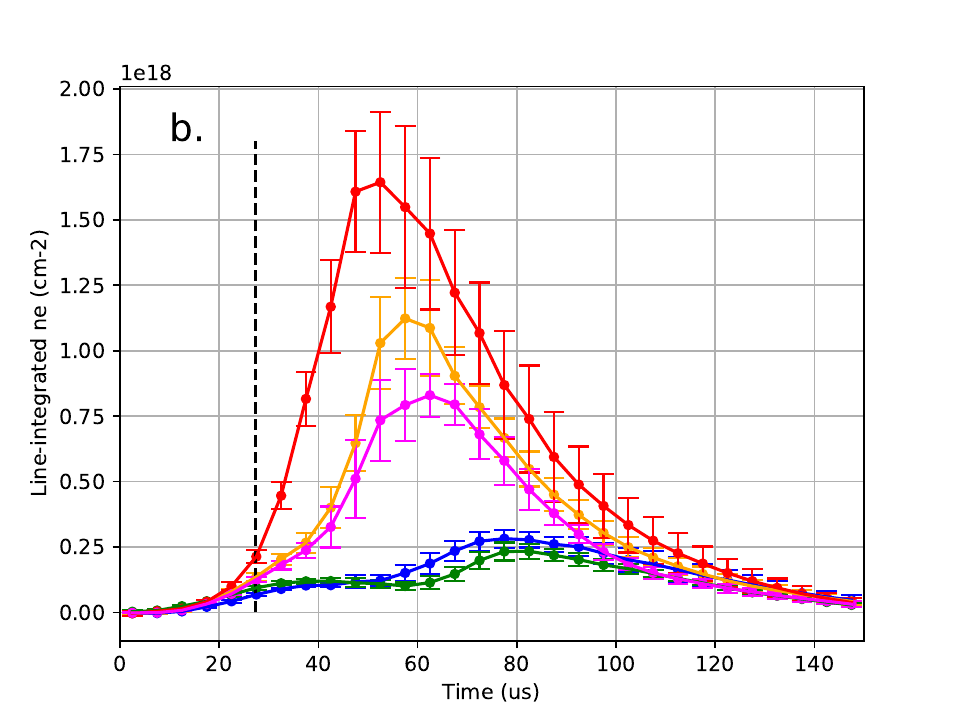}
\caption{\label{fig:_multishot_ne} Line-integrated plasma density profiles as measured by laser interferometry on a) single shot and b) average of $\sim$20 shots. Time zero is aligned to average launch time of the plasma jets, compensating for shot-to-shot jitter in trigger time.}
\end{figure}

\begin{figure}
\includegraphics[width=8.5cm]{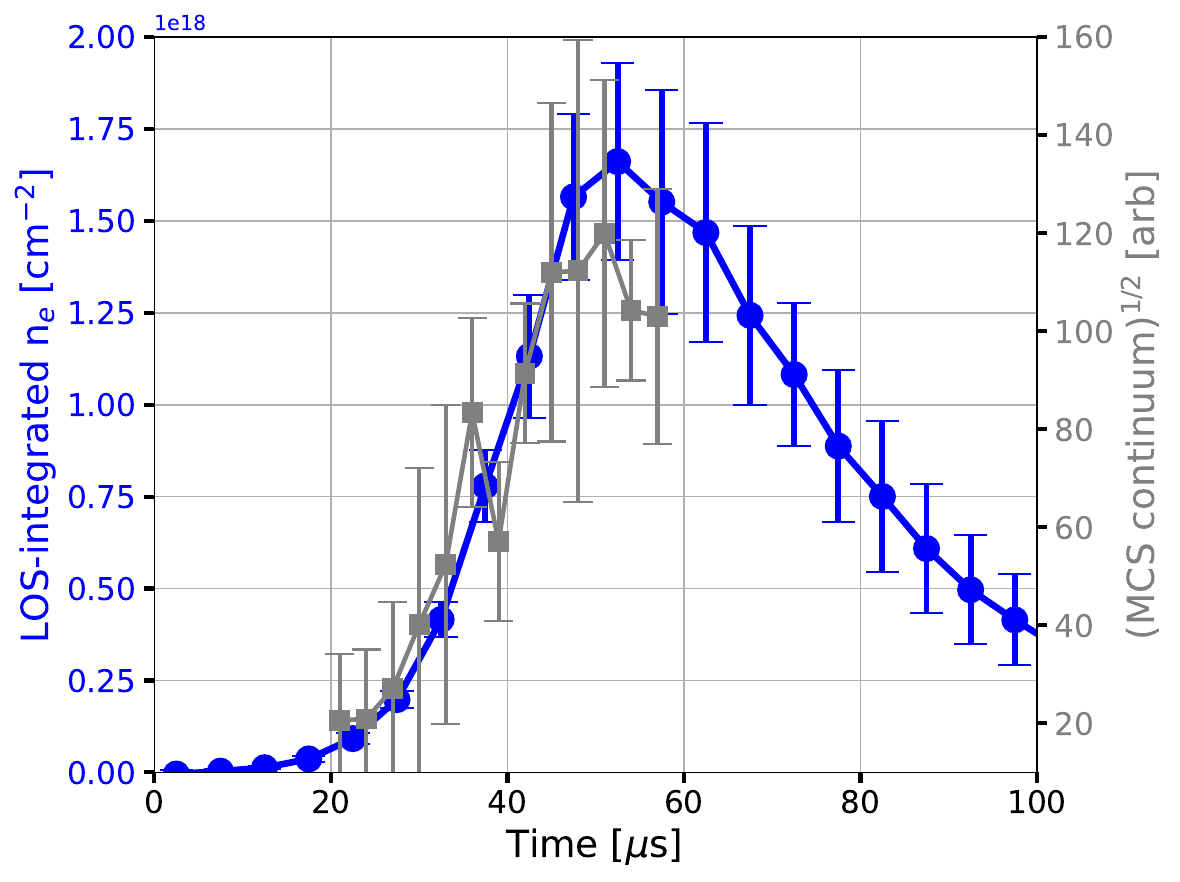}
\caption{\label{fig:Interferometry_ContinuumCompare} LOS-integrated electron density from interferometry (left axis) and MCS (right axis) both for their respective chord through chamber center. The latter is obtained as the square root of the MCS continuum in Fig. \ref{fig:MCS_ConsolidatedResults_cdeParts2}.}
\end{figure}

Inspection of the interferometry traces on both individual shots and multi-shot averages in Fig. \ref{fig:_multishot_ne} reveals several features. As the leading edge of the jets arrive, a slow increase of density is observed, preceding a sharp increase on the central chord. At later time, similar sharp increases are observed on the off-radial chords. In comparison with the experimental images in Fig. \ref{fig:HadlandImages2187_cmapMagma}, it is clear that the sharp increase is connected to the formation of the central expanding luminous region, which is the collisional stagnation of the liner and subsequent expansion / rebound of the stagnated liner plasma. The paired chords display decent symmetry within shot-to-shot variation on the early time rise of the density, with variations becoming observable on the rebound in both magnitude and timing of the rebound arrival. This is likely indicative of a small systematic asymmetry in the jet-to-jet delivered mass and pressure, leading to the observed asymmetry in the rebound over many shots. Spatial and temporal behavior of electron density determined from MCS continuum intensity, as shown in Fig. \ref{fig:Interferometry_ContinuumCompare}, are in general agreement with that determined via interferometry, and the time of peak density corresponds well with the bright stagnation time seen in Figs. \ref{fig:LinerMergeProgression_font2} and \ref{fig:HadlandImages2187_cmapMagma}.


\section{\label{sec:Discussion}Discussion\protect\\}

\subsection{Liner Uniformity}

A major question in PJMIF is whether a sufficiently smooth and uniform spherical liner can be formed from the merging of discrete hypersonic jets, without significant density perturbations from shock waves between the merging jets. The inertial fusion field in particular has encountered many manifestations of the Rayleigh-Taylor instability when attempting liner implosions, and found in all cases that smooth liners and shells of the utmost uniformity are essential to avoid enhanced instability growth.

In previous PLX studies of obliquely merging jets in pairs, trios, and clusters of 6 and 7 PLX plasma guns, inter-jet shocks along the merge plane of different jets were frequently observed, whereas it appears largely absent in the present experiment. This key difference is consistent with the former experiments typically exhibiting slower velocities and shallower mean merge angles between adjacent jets. In Ref[\onlinecite{myID_570}] it was found that when gun angular separations were increased from 11.6 to 20.5 degrees half-angle, interpentration length exceeded ion mean free path length and thus shock formation was substantially mitigated. In Ref[\onlinecite{myID_549}] significant improvements to jet-jet mass uniformity and modest improvements to velocity uniformity were implemented. Measurements of the merger of clusters of 6 or 7 jets indicated increased temperature and ionization, but could potentially still remain in the desired interpenetrating regime without shock formation.

In the present experiment, the mean separation angle $\bar{\alpha}$ between adjacent jets of 36 degrees constitutes the primary reason for not observing the inter-jet shocks observed in prior experiments, since their merge velocity $v_m = v_j sin(\bar{\alpha})$ is greater for larger angles. The characteristic ion-ion interpenetration length is given by:

\begin{equation}\label{eq:IonIonInterpenLength}
L_{ii,s} = \sum_{i'} \frac{v_i}{4\nu_{ii',s}}
\end{equation}

\noindent in which $\nu_{ii',s}$ is the slowing frequency through ion-ion collisions, given by:

\begin{equation}
\label{eq:IonIonSlowingFrequency}
\nu_{ii',s} = \num{9e-8}  \lambda_{ii'} \bar{Z}^2 \bar{Z}'^2 \Big( \frac{1}{\mu} + \frac{1}{\mu'} \Big) \frac{n_{i'} \mu^{1/2}}{\epsilon^{3/2}} 
\end{equation}

\noindent where $\lambda_{ii'}$ is the Coulomb logarithm between counterstreaming ions $i$ and $i'$, $\bar{Z}$ is the mean charge state, $\mu$ is the ion/proton mass ratio, $\epsilon$ is the energy associated with velocity $v$, in units of eV, and $n$ is density in cm\textsuperscript{-3}. If ions $i$ and $i'$ are of the same element (argon in this case), then the interpenetration length $L_{ii,s}$ scales as $v_i^4 / n_{i'}\bar{Z}^4$. As such, the velocity, merging angle, and charge state of the jets exert a large control over the degree of interpenetration.

If plasma-jet-driven liner implosions are scaled up to greatly increased plasma density, one can expect decreased jet interpenetration and re-emergence of collisional shock waves between the primary plasma jets. Due to the strong $v_i^4$ dependence of interpenetration length on velocity, increases in jet velocity concomitant with increases in liner density would have offsetting effects, resulting in the persistence of significant interpenetration even at many orders of magnitude greater energy densities. The exact termination point of this  scaling is not empirically known, but we conjecture that it will at maximum end with densities corresponding to the onset of optical thickness of the liner plasma, inhibiting effective radiative cooling of the liner and leading to an increase in liner average charge state $\bar{Z}$. Ultimately, this chain of events may lead to a relatively rapid onset of the liner collisionality occurring at a threshold in parameter space.  Future work may shed light on these and other detailed aspects of the jet merging and convergence phase and how they may be affected when scaling towards fusion-relevant conditions. The current experiment effectively provides a single encouraging data point showing that a liner free of primary shock waves can be assembled at these conditions.

\subsection{Comparison to Hydrodynamic Simulation}

\begin{figure}
\includegraphics[width=8.5cm]{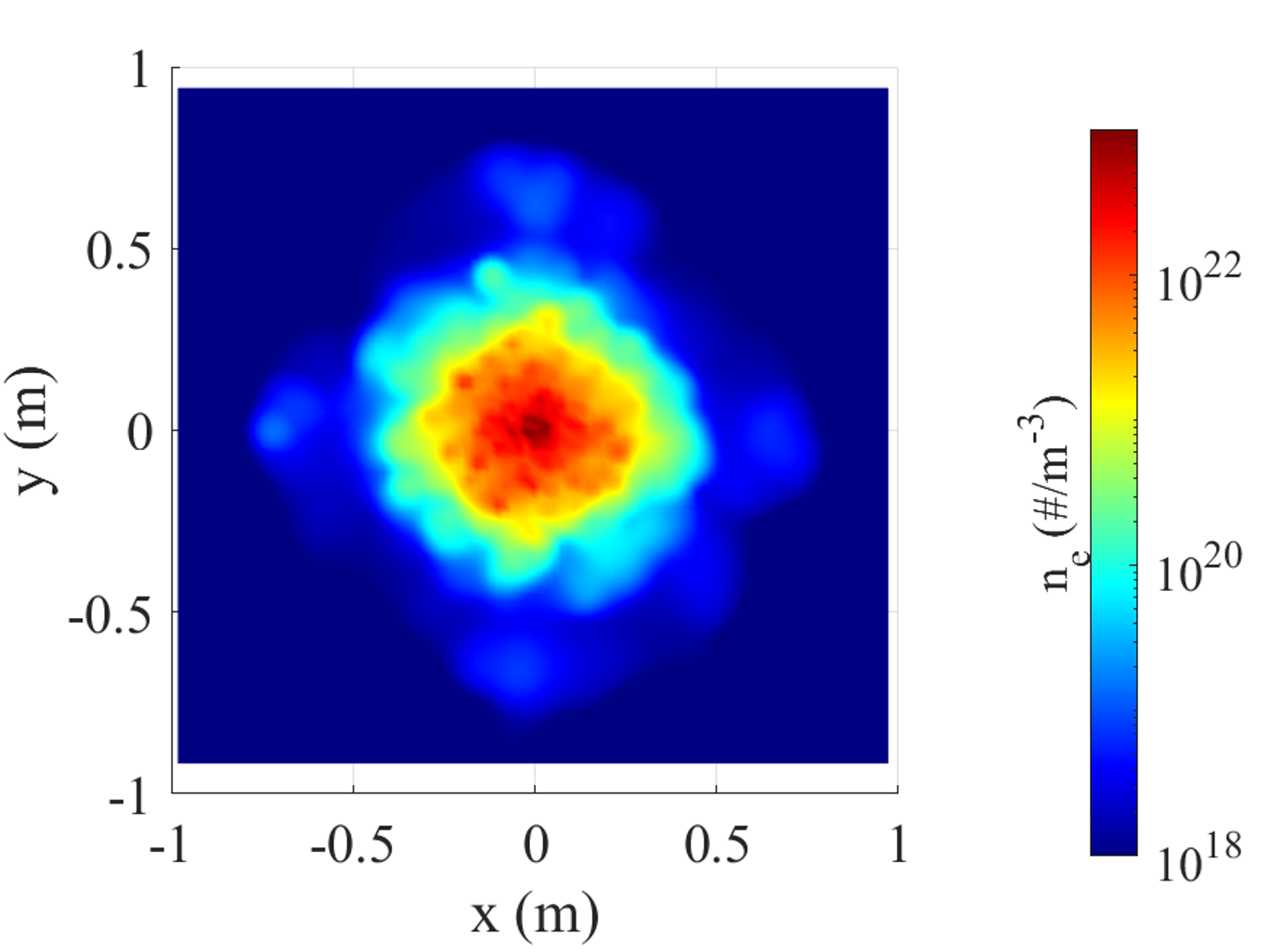}
\includegraphics[width=8.5cm]{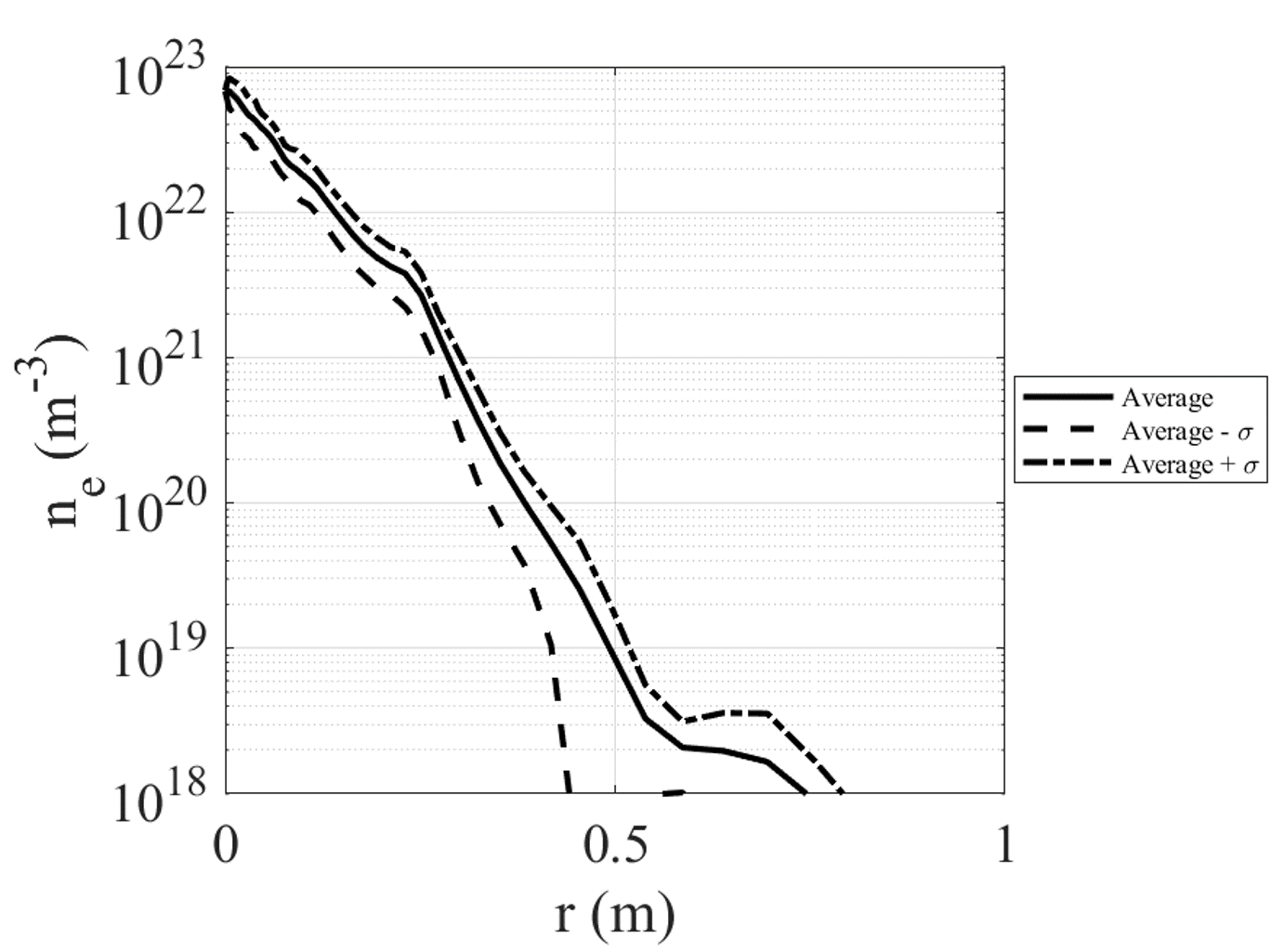}
\caption{\label{fig:simulation} Results from smooth-particle hydrodynamic simulation of the PLX 36-jet liner experiment using SPFMax code with ion slip model enabled. Observed spatial structure and radially-averaged plasma density at stagnation are in broad agreement with experimental observations (c.f. Figs. ~\ref{fig:LinerMergeProgression_font2},~\ref{fig:HiRes_Shot2170Analysis},~\ref{fig:_multishot_ne}.)}
\end{figure}

The experimental campaign is compared to smooth-particle hydrodynamics (SPH) simulations performed with the SPFMax code at UAH \cite{myID_581}. Simulations to date have been performed with uniform nominal jet initial conditions of density and velocity profile, not accounting for individual or shot-to-shot variations between jets. No effects of any residual magnetic field are included in the simulations (consistent with the cold dense jets being largely unmagnetized post-flight). An ion-slip model is used to account for the effects of kinetic interpenetration, leveraging the particle-based mechanics of the SPH code -- collisional interaction between fluid particles is adjusted by a multiplying factor based on equation~\ref{eq:IonIonSlowingFrequency}, with the effects that particles of high relative velocity experience little collisional interaction, and vice versa. Results of stagnation density profile with the ion-slip model turned on are shown in Figure~\ref{fig:simulation}, showing similar morphology to the experimental imaging and inferred density profiles. It is observed that simulation results with ion slip model turned on agree much more closely with the experimental results, re-enforcing the view that interpenetration is playing a strong role in high velocity liner formation.

\subsection{Liner Ram Pressure}

We can infer bounds of liner peak ram pressure ($\frac{1}{2} \rho v^2$) by combining the initial jet velocities measured from photodiodes ($V_{PD}$) for the upper bound velocity, and high-resolution ($V_{S}$) spectroscopy for lower bound velocity, plasma density $n_e$ from interferometry, and $\bar{Z}$ from MCS results to obtain $\rho = m_{Ar} (\frac{n_e}{\bar{Z}})$. To estimate $n_e$ from interferometry we take the peak line-integrated value of about $\num{2.0e18}$ cm\textsuperscript{-2}, over the apparent domain that imaging indicates to be brightest - around 15 cm. Thus the upper bound of ram pressure is about 123 bar, and the lower bound is about 20 bar. This level is consistent with modeling predictions that have incorporated the present jet lengths of $\sim30$ cm and moderate Mach numbers $\sim10-20$, revealing that there is a significant loss in liner density and ram pressure incurred due to the volumetric expansion of the individual jets en route to the liner merging radius, as well the extended length of the individual jets. If a thinner liner with sharper radial density profile can be formed, possibly by use of jets with higher Mach number and optimized launch velocity profile, the ram pressure has potential to be increased by many orders of magnitude at a given energy scale. Another path to increasing liner ram pressure and energy density is to simply decrease the liner standoff distance, effectively making the chamber smaller and starting the jets at a lesser radius. While this is attractive for the liner performance, it puts increased demands on the eventual system components to survive in a harsher fusion heat and radiation environment, and may complicate reactor engineering \cite{myID_571}.

\section{\label{sec:Conclusions}Conclusions\protect\\}

The present set of experiments at PLX delivers the most substantial experimental measurements to date to investigate the feasibility of the PJMIF liner formation approach. Spatial and temporal measurements are obtained of an argon liner formed via the merger of 36 discrete plasma jets, following the liner evolution through jet launch, merging, spherical convergence, stagnation, and rebound. Notably, conditions of moderate interpenetration between primary jets are observed to lead to the absence of dense primary shock structures in the early liner, allowing increased density uniformity to be achieved. If primary shock waves are mitigated in this way, it seems that arbitrarily smooth liners may be able to be formed with corresponding improvements in primary jet parameters and control. A clear transition from initial kinetic interpenetration to collisional interaction is observed, in accordance with theoretical expectation. The obtained diagnostic measurements also provide achieved absolute parameter values of ram pressure, stagnation density, and temperature, useful for benchmarking computational models and understanding the scaling of plasma liners to increased energy density. The results provide a first singular data point for these benchmarking purposes -- future experiments could seek to verify theoretical liner scalings by exploring the effect of perturbations to liner parameters.

\begin{acknowledgments}
The authors would particularly like to thank Tom Byvank, Edward Cruz, Levi Grantz, Ricardo Martinez, and many others who have contributed greatly to the current experiments and overall PLX project, as well as the pioneers of the PJMIF concept and PLX experiment, Francis Thio and Scott Hsu, respectively.

The information, data or work presented herein was funded in part by the Advanced Research Projects Agency-Energy (ARPA-E), U.S. Department of Energy under Award Numbers 20/CJ000/03/02 and DE-AR0001268, BETHE program. The views and opinions of authors expressed herein do not necessarily state or reflect those of the United States Government or any agency thereof.

This work was supported by the U.S. Department of Energy in part through the Los Alamos National Laboratory. Los Alamos National Laboratory is operated by Triad National Security, LLC, for the National Nuclear Security Administration of U.S. Department of Energy (Contract No. 89233218CNA000001).
\end{acknowledgments}


\section*{Data Availability}

The data that support the findings of this study are available from the corresponding author upon reasonable request.

\bibliography{LaJoie_Bibliography}

\begin{thebibliography}{32}%
\makeatletter
\providecommand \@ifxundefined [1]{%
 \@ifx{#1\undefined}
}%
\providecommand \@ifnum [1]{%
 \ifnum #1\expandafter \@firstoftwo
 \else \expandafter \@secondoftwo
 \fi
}%
\providecommand \@ifx [1]{%
 \ifx #1\expandafter \@firstoftwo
 \else \expandafter \@secondoftwo
 \fi
}%
\providecommand \natexlab [1]{#1}%
\providecommand \enquote  [1]{``#1''}%
\providecommand \bibnamefont  [1]{#1}%
\providecommand \bibfnamefont [1]{#1}%
\providecommand \citenamefont [1]{#1}%
\providecommand \href@noop [0]{\@secondoftwo}%
\providecommand \href [0]{\begingroup \@sanitize@url \@href}%
\providecommand \@href[1]{\@@startlink{#1}\@@href}%
\providecommand \@@href[1]{\endgroup#1\@@endlink}%
\providecommand \@sanitize@url [0]{\catcode `\\12\catcode `\$12\catcode
  `\&12\catcode `\#12\catcode `\^12\catcode `\_12\catcode `\%12\relax}%
\providecommand \@@startlink[1]{}%
\providecommand \@@endlink[0]{}%
\providecommand \url  [0]{\begingroup\@sanitize@url \@url }%
\providecommand \@url [1]{\endgroup\@href {#1}{\urlprefix }}%
\providecommand \urlprefix  [0]{URL }%
\providecommand \Eprint [0]{\href }%
\providecommand \doibase [0]{https://doi.org/}%
\providecommand \selectlanguage [0]{\@gobble}%
\providecommand \bibinfo  [0]{\@secondoftwo}%
\providecommand \bibfield  [0]{\@secondoftwo}%
\providecommand \translation [1]{[#1]}%
\providecommand \BibitemOpen [0]{}%
\providecommand \bibitemStop [0]{}%
\providecommand \bibitemNoStop [0]{.\EOS\space}%
\providecommand \EOS [0]{\spacefactor3000\relax}%
\providecommand \BibitemShut  [1]{\csname bibitem#1\endcsname}%
\let\auto@bib@innerbib\@empty
\bibitem [{\citenamefont {Kirkpatrick}, \citenamefont {Lindemuth},\ and\
  \citenamefont {Ward}(1995)}]{myID_577}%
  \BibitemOpen
  \bibfield  {author} {\bibinfo {author} {\bibfnamefont {R.~C.}\ \bibnamefont
  {Kirkpatrick}}, \bibinfo {author} {\bibfnamefont {I.~R.}\ \bibnamefont
  {Lindemuth}},\ and\ \bibinfo {author} {\bibfnamefont {M.~S.}\ \bibnamefont
  {Ward}},\ }\bibfield  {title} {\enquote {\bibinfo {title} {Magnetized target
  fusion: An overview},}\ }\href {https://doi.org/10.13182/FST95-A30382}
  {\bibfield  {journal} {\bibinfo  {journal} {Fusion Technology}\ }\textbf
  {\bibinfo {volume} {27}},\ \bibinfo {pages} {201--214} (\bibinfo {year}
  {1995})}\BibitemShut {NoStop}%
\bibitem [{\citenamefont {Lindemuth}(2015)}]{myID_578}%
  \BibitemOpen
  \bibfield  {author} {\bibinfo {author} {\bibfnamefont {I.~R.}\ \bibnamefont
  {Lindemuth}},\ }\bibfield  {title} {\enquote {\bibinfo {title} {The ignition
  design space of magnetized target fusion},}\ }\href
  {https://doi.org/10.1063/1.4937371} {\bibfield  {journal} {\bibinfo
  {journal} {Physics of Plasmas}\ }\textbf {\bibinfo {volume} {22}} (\bibinfo
  {year} {2015}),\ 10.1063/1.4937371}\BibitemShut {NoStop}%
\bibitem [{\citenamefont {Wurden}\ \emph {et~al.}(2016)\citenamefont {Wurden},
  \citenamefont {Hsu}, \citenamefont {Intrator}, \citenamefont {Grabowski},
  \citenamefont {Degnan}, \citenamefont {Domonkos}, \citenamefont {Turchi},
  \citenamefont {Campbell}, \citenamefont {Sinars}, \citenamefont {Herrmann},
  \citenamefont {Betti}, \citenamefont {Bauer}, \citenamefont {Lindemuth},
  \citenamefont {Siemon}, \citenamefont {Miller}, \citenamefont {Laberge},\
  and\ \citenamefont {Delage}}]{myID_579}%
  \BibitemOpen
  \bibfield  {author} {\bibinfo {author} {\bibfnamefont {G.~A.}\ \bibnamefont
  {Wurden}}, \bibinfo {author} {\bibfnamefont {S.~C.}\ \bibnamefont {Hsu}},
  \bibinfo {author} {\bibfnamefont {T.~P.}\ \bibnamefont {Intrator}}, \bibinfo
  {author} {\bibfnamefont {T.~C.}\ \bibnamefont {Grabowski}}, \bibinfo {author}
  {\bibfnamefont {J.~H.}\ \bibnamefont {Degnan}}, \bibinfo {author}
  {\bibfnamefont {M.}~\bibnamefont {Domonkos}}, \bibinfo {author}
  {\bibfnamefont {P.~J.}\ \bibnamefont {Turchi}}, \bibinfo {author}
  {\bibfnamefont {E.~M.}\ \bibnamefont {Campbell}}, \bibinfo {author}
  {\bibfnamefont {D.~B.}\ \bibnamefont {Sinars}}, \bibinfo {author}
  {\bibfnamefont {M.~C.}\ \bibnamefont {Herrmann}}, \bibinfo {author}
  {\bibfnamefont {R.}~\bibnamefont {Betti}}, \bibinfo {author} {\bibfnamefont
  {B.~S.}\ \bibnamefont {Bauer}}, \bibinfo {author} {\bibfnamefont {I.~R.}\
  \bibnamefont {Lindemuth}}, \bibinfo {author} {\bibfnamefont {R.~E.}\
  \bibnamefont {Siemon}}, \bibinfo {author} {\bibfnamefont {R.~L.}\
  \bibnamefont {Miller}}, \bibinfo {author} {\bibfnamefont {M.}~\bibnamefont
  {Laberge}},\ and\ \bibinfo {author} {\bibfnamefont {M.}~\bibnamefont
  {Delage}},\ }\bibfield  {title} {\enquote {\bibinfo {title} {Magneto-inertial
  fusion},}\ }\href {https://doi.org/10.1007/s10894-015-0038-x} {\bibfield
  {journal} {\bibinfo  {journal} {Journal of Fusion Energy}\ }\textbf {\bibinfo
  {volume} {35}},\ \bibinfo {pages} {69–77} (\bibinfo {year}
  {2016})}\BibitemShut {NoStop}%
\bibitem [{\citenamefont {Hsu}\ \emph {et~al.}(2018)\citenamefont {Hsu},
  \citenamefont {Langendorf}, \citenamefont {Yates}, \citenamefont {Dunn},
  \citenamefont {Brockington}, \citenamefont {Case}, \citenamefont {Cruz},
  \citenamefont {Witherspoon}, \citenamefont {Gilmore}, \citenamefont
  {Cassibry}, \citenamefont {Samulyak}, \citenamefont {Stoltz}, \citenamefont
  {Schillo}, \citenamefont {Shih}, \citenamefont {Beckwith},\ and\
  \citenamefont {Thio}}]{myID_563}%
  \BibitemOpen
  \bibfield  {author} {\bibinfo {author} {\bibfnamefont {S.~C.}\ \bibnamefont
  {Hsu}}, \bibinfo {author} {\bibfnamefont {S.~J.}\ \bibnamefont {Langendorf}},
  \bibinfo {author} {\bibfnamefont {K.~C.}\ \bibnamefont {Yates}}, \bibinfo
  {author} {\bibfnamefont {J.~P.}\ \bibnamefont {Dunn}}, \bibinfo {author}
  {\bibfnamefont {S.}~\bibnamefont {Brockington}}, \bibinfo {author}
  {\bibfnamefont {A.}~\bibnamefont {Case}}, \bibinfo {author} {\bibfnamefont
  {E.}~\bibnamefont {Cruz}}, \bibinfo {author} {\bibfnamefont {F.~D.}\
  \bibnamefont {Witherspoon}}, \bibinfo {author} {\bibfnamefont {M.~A.}\
  \bibnamefont {Gilmore}}, \bibinfo {author} {\bibfnamefont {J.~T.}\
  \bibnamefont {Cassibry}}, \bibinfo {author} {\bibfnamefont {R.}~\bibnamefont
  {Samulyak}}, \bibinfo {author} {\bibfnamefont {P.}~\bibnamefont {Stoltz}},
  \bibinfo {author} {\bibfnamefont {K.}~\bibnamefont {Schillo}}, \bibinfo
  {author} {\bibfnamefont {W.}~\bibnamefont {Shih}}, \bibinfo {author}
  {\bibfnamefont {K.}~\bibnamefont {Beckwith}},\ and\ \bibinfo {author}
  {\bibfnamefont {Y.~C.~F.}\ \bibnamefont {Thio}},\ }\bibfield  {title}
  {\enquote {\bibinfo {title} {Experiment to form and characterize a section of
  a spherically imploding plasma liner},}\ }\href
  {https://doi.org/10.1109/TPS.2017.2779421} {\bibfield  {journal} {\bibinfo
  {journal} {IEEE Transactions on Plasma Science}\ }\textbf {\bibinfo {volume}
  {46}},\ \bibinfo {pages} {1951--1961} (\bibinfo {year} {2018})}\BibitemShut
  {NoStop}%
\bibitem [{\citenamefont {Thio}\ \emph {et~al.}(2019)\citenamefont {Thio},
  \citenamefont {Hsu}, \citenamefont {Witherspoon}, \citenamefont {Cruz},
  \citenamefont {Case}, \citenamefont {Langendorf}, \citenamefont {Yates},
  \citenamefont {Dunn}, \citenamefont {Cassibry}, \citenamefont {Samulyak},
  \citenamefont {Stoltz}, \citenamefont {Brockington}, \citenamefont
  {Williams}, \citenamefont {Luna}, \citenamefont {Becker},\ and\ \citenamefont
  {Cook}}]{myID_520}%
  \BibitemOpen
  \bibfield  {author} {\bibinfo {author} {\bibfnamefont {Y.~C.~F.}\
  \bibnamefont {Thio}}, \bibinfo {author} {\bibfnamefont {S.~C.}\ \bibnamefont
  {Hsu}}, \bibinfo {author} {\bibfnamefont {F.~D.}\ \bibnamefont
  {Witherspoon}}, \bibinfo {author} {\bibfnamefont {E.}~\bibnamefont {Cruz}},
  \bibinfo {author} {\bibfnamefont {A.}~\bibnamefont {Case}}, \bibinfo {author}
  {\bibfnamefont {S.}~\bibnamefont {Langendorf}}, \bibinfo {author}
  {\bibfnamefont {K.}~\bibnamefont {Yates}}, \bibinfo {author} {\bibfnamefont
  {J.}~\bibnamefont {Dunn}}, \bibinfo {author} {\bibfnamefont {J.}~\bibnamefont
  {Cassibry}}, \bibinfo {author} {\bibfnamefont {R.}~\bibnamefont {Samulyak}},
  \bibinfo {author} {\bibfnamefont {P.}~\bibnamefont {Stoltz}}, \bibinfo
  {author} {\bibfnamefont {S.~J.}\ \bibnamefont {Brockington}}, \bibinfo
  {author} {\bibfnamefont {A.}~\bibnamefont {Williams}}, \bibinfo {author}
  {\bibfnamefont {M.}~\bibnamefont {Luna}}, \bibinfo {author} {\bibfnamefont
  {R.}~\bibnamefont {Becker}},\ and\ \bibinfo {author} {\bibfnamefont
  {A.}~\bibnamefont {Cook}},\ }\bibfield  {title} {\enquote {\bibinfo {title}
  {Plasma-jet-driven magneto-inertial fusion},}\ }\href
  {https://doi.org/10.1080/15361055.2019.1598736} {\bibfield  {journal}
  {\bibinfo  {journal} {Fusion Science and Technology}\ }\textbf {\bibinfo
  {volume} {75}},\ \bibinfo {pages} {581–598} (\bibinfo {year}
  {2019})}\BibitemShut {NoStop}%
\bibitem [{\citenamefont {Yager-Elorriaga}\ \emph {et~al.}(2022)\citenamefont
  {Yager-Elorriaga}, \citenamefont {Gomez}, \citenamefont {Ruiz}, \citenamefont
  {Slutz}, \citenamefont {Harvey-Thompson}, \citenamefont {Jennings},
  \citenamefont {Knapp}, \citenamefont {Schmit}, \citenamefont {Weis},
  \citenamefont {Awe}, \citenamefont {Chandler}, \citenamefont {Mangan},
  \citenamefont {Myers}, \citenamefont {Fein}, \citenamefont {Galloway},
  \citenamefont {Geissel}, \citenamefont {Glinsky}, \citenamefont {Hansen},
  \citenamefont {Harding}, \citenamefont {Lamppa}, \citenamefont {Lewis},
  \citenamefont {Rambo}, \citenamefont {Robertson}, \citenamefont {Savage},
  \citenamefont {Shipley}, \citenamefont {Smith}, \citenamefont {Schwarz},
  \citenamefont {Ampleford}, \citenamefont {Beckwith}, \citenamefont
  {Peterson}, \citenamefont {Porter}, \citenamefont {Rochau},\ and\
  \citenamefont {Sinars}}]{myID_566}%
  \BibitemOpen
  \bibfield  {author} {\bibinfo {author} {\bibfnamefont {D.}~\bibnamefont
  {Yager-Elorriaga}}, \bibinfo {author} {\bibfnamefont {M.}~\bibnamefont
  {Gomez}}, \bibinfo {author} {\bibfnamefont {D.}~\bibnamefont {Ruiz}},
  \bibinfo {author} {\bibfnamefont {S.}~\bibnamefont {Slutz}}, \bibinfo
  {author} {\bibfnamefont {A.}~\bibnamefont {Harvey-Thompson}}, \bibinfo
  {author} {\bibfnamefont {C.}~\bibnamefont {Jennings}}, \bibinfo {author}
  {\bibfnamefont {P.}~\bibnamefont {Knapp}}, \bibinfo {author} {\bibfnamefont
  {P.}~\bibnamefont {Schmit}}, \bibinfo {author} {\bibfnamefont
  {M.}~\bibnamefont {Weis}}, \bibinfo {author} {\bibfnamefont {T.}~\bibnamefont
  {Awe}}, \bibinfo {author} {\bibfnamefont {G.}~\bibnamefont {Chandler}},
  \bibinfo {author} {\bibfnamefont {M.}~\bibnamefont {Mangan}}, \bibinfo
  {author} {\bibfnamefont {C.}~\bibnamefont {Myers}}, \bibinfo {author}
  {\bibfnamefont {J.}~\bibnamefont {Fein}}, \bibinfo {author} {\bibfnamefont
  {B.}~\bibnamefont {Galloway}}, \bibinfo {author} {\bibfnamefont
  {M.}~\bibnamefont {Geissel}}, \bibinfo {author} {\bibfnamefont
  {M.}~\bibnamefont {Glinsky}}, \bibinfo {author} {\bibfnamefont
  {S.}~\bibnamefont {Hansen}}, \bibinfo {author} {\bibfnamefont
  {E.}~\bibnamefont {Harding}}, \bibinfo {author} {\bibfnamefont
  {D.}~\bibnamefont {Lamppa}}, \bibinfo {author} {\bibfnamefont
  {W.}~\bibnamefont {Lewis}}, \bibinfo {author} {\bibfnamefont
  {P.}~\bibnamefont {Rambo}}, \bibinfo {author} {\bibfnamefont
  {G.}~\bibnamefont {Robertson}}, \bibinfo {author} {\bibfnamefont
  {M.}~\bibnamefont {Savage}}, \bibinfo {author} {\bibfnamefont
  {G.}~\bibnamefont {Shipley}}, \bibinfo {author} {\bibfnamefont
  {I.}~\bibnamefont {Smith}}, \bibinfo {author} {\bibfnamefont
  {J.}~\bibnamefont {Schwarz}}, \bibinfo {author} {\bibfnamefont
  {D.}~\bibnamefont {Ampleford}}, \bibinfo {author} {\bibfnamefont
  {K.}~\bibnamefont {Beckwith}}, \bibinfo {author} {\bibfnamefont
  {K.}~\bibnamefont {Peterson}}, \bibinfo {author} {\bibfnamefont
  {J.}~\bibnamefont {Porter}}, \bibinfo {author} {\bibfnamefont
  {G.}~\bibnamefont {Rochau}},\ and\ \bibinfo {author} {\bibfnamefont
  {D.}~\bibnamefont {Sinars}},\ }\bibfield  {title} {\enquote {\bibinfo {title}
  {An overview of magneto-inertial fusion on the z machine at sandia national
  laboratories},}\ }\href {https://doi.org/10.1088/1741-4326/ac2dbe} {\bibfield
   {journal} {\bibinfo  {journal} {Nuclear Fusion}\ }\textbf {\bibinfo {volume}
  {62}} (\bibinfo {year} {2022}),\ 10.1088/1741-4326/ac2dbe}\BibitemShut
  {NoStop}%
\bibitem [{\citenamefont {Zhang}\ \emph {et~al.}(2019)\citenamefont {Zhang},
  \citenamefont {Shumlak}, \citenamefont {Nelson}, \citenamefont {Golingo},
  \citenamefont {Weber}, \citenamefont {Stepanov}, \citenamefont {Claveau},
  \citenamefont {Forbes}, \citenamefont {Draper}, \citenamefont {Mitrani},
  \citenamefont {McLean}, \citenamefont {Tummel}, \citenamefont {Higginson},\
  and\ \citenamefont {Cooper}}]{myID_567}%
  \BibitemOpen
  \bibfield  {author} {\bibinfo {author} {\bibfnamefont {Y.}~\bibnamefont
  {Zhang}}, \bibinfo {author} {\bibfnamefont {U.}~\bibnamefont {Shumlak}},
  \bibinfo {author} {\bibfnamefont {B.~A.}\ \bibnamefont {Nelson}}, \bibinfo
  {author} {\bibfnamefont {R.~P.}\ \bibnamefont {Golingo}}, \bibinfo {author}
  {\bibfnamefont {T.~R.}\ \bibnamefont {Weber}}, \bibinfo {author}
  {\bibfnamefont {A.~D.}\ \bibnamefont {Stepanov}}, \bibinfo {author}
  {\bibfnamefont {E.~L.}\ \bibnamefont {Claveau}}, \bibinfo {author}
  {\bibfnamefont {E.~G.}\ \bibnamefont {Forbes}}, \bibinfo {author}
  {\bibfnamefont {Z.~T.}\ \bibnamefont {Draper}}, \bibinfo {author}
  {\bibfnamefont {J.~M.}\ \bibnamefont {Mitrani}}, \bibinfo {author}
  {\bibfnamefont {H.~S.}\ \bibnamefont {McLean}}, \bibinfo {author}
  {\bibfnamefont {K.~K.}\ \bibnamefont {Tummel}}, \bibinfo {author}
  {\bibfnamefont {D.~P.}\ \bibnamefont {Higginson}},\ and\ \bibinfo {author}
  {\bibfnamefont {C.~M.}\ \bibnamefont {Cooper}},\ }\bibfield  {title}
  {\enquote {\bibinfo {title} {Sustained neutron production from a sheared-flow
  stabilized z pinch},}\ }\href
  {https://doi.org/10.1103/PhysRevLett.122.135001} {\bibfield  {journal}
  {\bibinfo  {journal} {Physical Review Letters}\ }\textbf {\bibinfo {volume}
  {122}} (\bibinfo {year} {2019}),\ 10.1103/PhysRevLett.122.135001}\BibitemShut
  {NoStop}%
\bibitem [{\citenamefont {O'Shea}\ \emph {et~al.}(2018)\citenamefont {O'Shea},
  \citenamefont {Laberge}, \citenamefont {Donaldson}, \citenamefont {Delage},
  \citenamefont {Mossman}, \citenamefont {Reynolds}, \citenamefont
  {de~Vietien},\ and\ \citenamefont {Team}}]{myID_572}%
  \BibitemOpen
  \bibfield  {author} {\bibinfo {author} {\bibfnamefont {P.}~\bibnamefont
  {O'Shea}}, \bibinfo {author} {\bibfnamefont {M.}~\bibnamefont {Laberge}},
  \bibinfo {author} {\bibfnamefont {M.}~\bibnamefont {Donaldson}}, \bibinfo
  {author} {\bibfnamefont {M.}~\bibnamefont {Delage}}, \bibinfo {author}
  {\bibfnamefont {A.}~\bibnamefont {Mossman}}, \bibinfo {author} {\bibfnamefont
  {M.}~\bibnamefont {Reynolds}}, \bibinfo {author} {\bibfnamefont
  {P.}~\bibnamefont {de~Vietien}},\ and\ \bibinfo {author} {\bibfnamefont
  {G.~F.}\ \bibnamefont {Team}},\ }\bibfield  {title} {\enquote {\bibinfo
  {title} {Magnetized target fusion at general fusion: An overview},}\ }in\
  \href@noop {} {\emph {\bibinfo {booktitle} {60th Annual Meeting of the APS
  Division of Plasma Physics, Portland, Oregon, USA}}}\ (\bibinfo {year}
  {2018})\ pp.\ \bibinfo {pages} {5--9}\BibitemShut {NoStop}%
\bibitem [{\citenamefont {Turchi}\ \emph {et~al.}(1976)\citenamefont {Turchi},
  \citenamefont {Cooper}, \citenamefont {Ford},\ and\ \citenamefont
  {Jenkins}}]{myID_573}%
  \BibitemOpen
  \bibfield  {author} {\bibinfo {author} {\bibfnamefont {P.~J.}\ \bibnamefont
  {Turchi}}, \bibinfo {author} {\bibfnamefont {A.~L.}\ \bibnamefont {Cooper}},
  \bibinfo {author} {\bibfnamefont {R.}~\bibnamefont {Ford}},\ and\ \bibinfo
  {author} {\bibfnamefont {D.~J.}\ \bibnamefont {Jenkins}},\ }\bibfield
  {title} {\enquote {\bibinfo {title} {Rotational stabilization of an imploding
  liquid cylinder},}\ }\href@noop {} {\bibfield  {journal} {\bibinfo  {journal}
  {Physical Review Letters}\ }\textbf {\bibinfo {volume} {36}},\ \bibinfo
  {pages} {1546} (\bibinfo {year} {1976})}\BibitemShut {NoStop}%
\bibitem [{\citenamefont {Wessel}\ \emph {et~al.}(2015)\citenamefont {Wessel},
  \citenamefont {Ney}, \citenamefont {Presura} \emph {et~al.}}]{myID_574}%
  \BibitemOpen
  \bibfield  {author} {\bibinfo {author} {\bibfnamefont {F.~J.}\ \bibnamefont
  {Wessel}}, \bibinfo {author} {\bibfnamefont {P.}~\bibnamefont {Ney}},
  \bibinfo {author} {\bibfnamefont {R.}~\bibnamefont {Presura}}, \emph
  {et~al.},\ }\bibfield  {title} {\enquote {\bibinfo {title} {Fusion in a
  staged $ z $-pinch},}\ }\href@noop {} {\bibfield  {journal} {\bibinfo
  {journal} {IEEE Transactions on Plasma Science}\ }\textbf {\bibinfo {volume}
  {43}},\ \bibinfo {pages} {2463--2468} (\bibinfo {year} {2015})}\BibitemShut
  {NoStop}%
\bibitem [{\citenamefont {Thio}\ \emph {et~al.}(2001)\citenamefont {Thio},
  \citenamefont {Knapp}, \citenamefont {Kirkpatrick}, \citenamefont {Siemon},\
  and\ \citenamefont {Turchi}}]{myID_545}%
  \BibitemOpen
  \bibfield  {author} {\bibinfo {author} {\bibfnamefont {Y.~C.~F.}\
  \bibnamefont {Thio}}, \bibinfo {author} {\bibfnamefont {C.~E.}\ \bibnamefont
  {Knapp}}, \bibinfo {author} {\bibfnamefont {R.~C.}\ \bibnamefont
  {Kirkpatrick}}, \bibinfo {author} {\bibfnamefont {R.~E.}\ \bibnamefont
  {Siemon}},\ and\ \bibinfo {author} {\bibfnamefont {P.~J.}\ \bibnamefont
  {Turchi}},\ }\bibfield  {title} {\enquote {\bibinfo {title} {A physics
  exploratory experiment on plasma liner formation},}\ }\href
  {https://doi.org/10.1023/A:1019813528507} {\bibfield  {journal} {\bibinfo
  {journal} {Journal of Fusion Energy}\ }\textbf {\bibinfo {volume} {20}},\
  \bibinfo {pages} {1--11} (\bibinfo {year} {2001})}\BibitemShut {NoStop}%
\bibitem [{\citenamefont {Hsu}\ \emph {et~al.}(2012{\natexlab{a}})\citenamefont
  {Hsu}, \citenamefont {Awe}, \citenamefont {Brockington}, \citenamefont
  {Case}, \citenamefont {Cassibry}, \citenamefont {Kagan}, \citenamefont
  {Messer}, \citenamefont {Stanic}, \citenamefont {Tang}, \citenamefont
  {Welch},\ and\ \citenamefont {Witherspoon}}]{myID_519}%
  \BibitemOpen
  \bibfield  {author} {\bibinfo {author} {\bibfnamefont {S.~C.}\ \bibnamefont
  {Hsu}}, \bibinfo {author} {\bibfnamefont {T.~J.}\ \bibnamefont {Awe}},
  \bibinfo {author} {\bibfnamefont {S.}~\bibnamefont {Brockington}}, \bibinfo
  {author} {\bibfnamefont {A.}~\bibnamefont {Case}}, \bibinfo {author}
  {\bibfnamefont {J.~T.}\ \bibnamefont {Cassibry}}, \bibinfo {author}
  {\bibfnamefont {G.}~\bibnamefont {Kagan}}, \bibinfo {author} {\bibfnamefont
  {S.~J.}\ \bibnamefont {Messer}}, \bibinfo {author} {\bibfnamefont
  {M.}~\bibnamefont {Stanic}}, \bibinfo {author} {\bibfnamefont
  {X.}~\bibnamefont {Tang}}, \bibinfo {author} {\bibfnamefont {D.~R.}\
  \bibnamefont {Welch}},\ and\ \bibinfo {author} {\bibfnamefont {F.~D.}\
  \bibnamefont {Witherspoon}},\ }\bibfield  {title} {\enquote {\bibinfo {title}
  {Spherically imploding plasma liners as a standoff driver for magnetoinertial
  fusion},}\ }\href {https://doi.org/10.1109/TPS.2012.2186829} {\bibfield
  {journal} {\bibinfo  {journal} {IEEE Transactions on Plasma Science}\
  }\textbf {\bibinfo {volume} {40}},\ \bibinfo {pages} {1287--1298} (\bibinfo
  {year} {2012}{\natexlab{a}})}\BibitemShut {NoStop}%
\bibitem [{\citenamefont {Lindl}(1995)}]{myID_534}%
  \BibitemOpen
  \bibfield  {author} {\bibinfo {author} {\bibfnamefont {J.}~\bibnamefont
  {Lindl}},\ }\bibfield  {title} {\enquote {\bibinfo {title} {Development of
  the indirect-drive approach to inertial confinement fusion and the target
  physics basis for ignition and gain},}\ }\href
  {https://doi.org/10.1063/1.871025} {\bibfield  {journal} {\bibinfo  {journal}
  {Physics of Plasmas}\ }\textbf {\bibinfo {volume} {2}},\ \bibinfo {pages}
  {3933--4024} (\bibinfo {year} {1995})}\BibitemShut {NoStop}%
\bibitem [{\citenamefont {Craxton}\ \emph {et~al.}(2015)\citenamefont
  {Craxton}, \citenamefont {Anderson}, \citenamefont {Boehly}, \citenamefont
  {Goncharov}, \citenamefont {Harding}, \citenamefont {Knauer}, \citenamefont
  {McCrory}, \citenamefont {McKenty}, \citenamefont {Meyerhofer}, \citenamefont
  {Myatt}, \citenamefont {Schmitt}, \citenamefont {Sethian}, \citenamefont
  {Short}, \citenamefont {Skupsky}, \citenamefont {Theobald}, \citenamefont
  {Kruer}, \citenamefont {Tanaka}, \citenamefont {Betti}, \citenamefont
  {Collins}, \citenamefont {Delettrez}, \citenamefont {Hu}, \citenamefont
  {Marozas}, \citenamefont {Maximov}, \citenamefont {Michel}, \citenamefont
  {Radha}, \citenamefont {Regan}, \citenamefont {Sangster}, \citenamefont
  {Seka}, \citenamefont {Solodov}, \citenamefont {Soures}, \citenamefont
  {Stoeckl},\ and\ \citenamefont {Zuegel}}]{myID_522}%
  \BibitemOpen
  \bibfield  {author} {\bibinfo {author} {\bibfnamefont {R.~S.}\ \bibnamefont
  {Craxton}}, \bibinfo {author} {\bibfnamefont {K.~S.}\ \bibnamefont
  {Anderson}}, \bibinfo {author} {\bibfnamefont {T.~R.}\ \bibnamefont
  {Boehly}}, \bibinfo {author} {\bibfnamefont {V.~N.}\ \bibnamefont
  {Goncharov}}, \bibinfo {author} {\bibfnamefont {D.~R.}\ \bibnamefont
  {Harding}}, \bibinfo {author} {\bibfnamefont {J.~P.}\ \bibnamefont {Knauer}},
  \bibinfo {author} {\bibfnamefont {R.~L.}\ \bibnamefont {McCrory}}, \bibinfo
  {author} {\bibfnamefont {P.~W.}\ \bibnamefont {McKenty}}, \bibinfo {author}
  {\bibfnamefont {D.~D.}\ \bibnamefont {Meyerhofer}}, \bibinfo {author}
  {\bibfnamefont {J.~F.}\ \bibnamefont {Myatt}}, \bibinfo {author}
  {\bibfnamefont {A.~J.}\ \bibnamefont {Schmitt}}, \bibinfo {author}
  {\bibfnamefont {J.~D.}\ \bibnamefont {Sethian}}, \bibinfo {author}
  {\bibfnamefont {R.~W.}\ \bibnamefont {Short}}, \bibinfo {author}
  {\bibfnamefont {S.}~\bibnamefont {Skupsky}}, \bibinfo {author} {\bibfnamefont
  {W.}~\bibnamefont {Theobald}}, \bibinfo {author} {\bibfnamefont {W.~L.}\
  \bibnamefont {Kruer}}, \bibinfo {author} {\bibfnamefont {K.}~\bibnamefont
  {Tanaka}}, \bibinfo {author} {\bibfnamefont {R.}~\bibnamefont {Betti}},
  \bibinfo {author} {\bibfnamefont {T.~J.~B.}\ \bibnamefont {Collins}},
  \bibinfo {author} {\bibfnamefont {J.~A.}\ \bibnamefont {Delettrez}}, \bibinfo
  {author} {\bibfnamefont {S.~X.}\ \bibnamefont {Hu}}, \bibinfo {author}
  {\bibfnamefont {J.~A.}\ \bibnamefont {Marozas}}, \bibinfo {author}
  {\bibfnamefont {A.~V.}\ \bibnamefont {Maximov}}, \bibinfo {author}
  {\bibfnamefont {D.~T.}\ \bibnamefont {Michel}}, \bibinfo {author}
  {\bibfnamefont {P.~B.}\ \bibnamefont {Radha}}, \bibinfo {author}
  {\bibfnamefont {S.~P.}\ \bibnamefont {Regan}}, \bibinfo {author}
  {\bibfnamefont {T.~C.}\ \bibnamefont {Sangster}}, \bibinfo {author}
  {\bibfnamefont {W.}~\bibnamefont {Seka}}, \bibinfo {author} {\bibfnamefont
  {A.~A.}\ \bibnamefont {Solodov}}, \bibinfo {author} {\bibfnamefont {J.~M.}\
  \bibnamefont {Soures}}, \bibinfo {author} {\bibfnamefont {C.}~\bibnamefont
  {Stoeckl}},\ and\ \bibinfo {author} {\bibfnamefont {J.~D.}\ \bibnamefont
  {Zuegel}},\ }\bibfield  {title} {\enquote {\bibinfo {title} {Direct-drive
  inertial confinement fusion: A review},}\ }\href
  {https://doi.org/10.1063/1.4934714} {\bibfield  {journal} {\bibinfo
  {journal} {Physics of Plasmas}\ }\textbf {\bibinfo {volume} {22}},\ \bibinfo
  {pages} {110501} (\bibinfo {year} {2015})}\BibitemShut {NoStop}%
\bibitem [{\citenamefont {Karino}\ \emph {et~al.}(2016)\citenamefont {Karino},
  \citenamefont {Kawata}, \citenamefont {Kondo}, \citenamefont {IInuma},
  \citenamefont {Kubo}, \citenamefont {Kato},\ and\ \citenamefont
  {Ogoyski}}]{myID_521}%
  \BibitemOpen
  \bibfield  {author} {\bibinfo {author} {\bibfnamefont {T.}~\bibnamefont
  {Karino}}, \bibinfo {author} {\bibfnamefont {S.}~\bibnamefont {Kawata}},
  \bibinfo {author} {\bibfnamefont {S.}~\bibnamefont {Kondo}}, \bibinfo
  {author} {\bibfnamefont {T.}~\bibnamefont {IInuma}}, \bibinfo {author}
  {\bibfnamefont {T.}~\bibnamefont {Kubo}}, \bibinfo {author} {\bibfnamefont
  {H.}~\bibnamefont {Kato}},\ and\ \bibinfo {author} {\bibfnamefont {A.~I.}\
  \bibnamefont {Ogoyski}},\ }\bibfield  {title} {\enquote {\bibinfo {title}
  {Target implosion uniformity in heavy-ion fusion},}\ }\href
  {https://doi.org/10.1017/S0263034616000690} {\bibfield  {journal} {\bibinfo
  {journal} {Laser and Particle Beams}\ }\textbf {\bibinfo {volume} {34}},\
  \bibinfo {pages} {735–741} (\bibinfo {year} {2016})}\BibitemShut {NoStop}%
\bibitem [{\citenamefont {Hsu}\ and\ \citenamefont
  {Langendorf}(2019)}]{myID_565}%
  \BibitemOpen
  \bibfield  {author} {\bibinfo {author} {\bibfnamefont {S.~C.}\ \bibnamefont
  {Hsu}}\ and\ \bibinfo {author} {\bibfnamefont {S.~J.}\ \bibnamefont
  {Langendorf}},\ }\bibfield  {title} {\enquote {\bibinfo {title} {Magnetized
  plasma target for plasma-jet-driven magneto-inertial fusion},}\ }\href
  {https://doi.org/10.1007/s10894-018-0168-z} {\bibfield  {journal} {\bibinfo
  {journal} {Journal of Fusion Energy}\ }\textbf {\bibinfo {volume} {38}},\
  \bibinfo {pages} {182–198} (\bibinfo {year} {2019})}\BibitemShut {NoStop}%
\bibitem [{\citenamefont {Knapp}\ and\ \citenamefont
  {Kirkpatrick}(2014)}]{myID_575}%
  \BibitemOpen
  \bibfield  {author} {\bibinfo {author} {\bibfnamefont {C.~E.}\ \bibnamefont
  {Knapp}}\ and\ \bibinfo {author} {\bibfnamefont {R.~C.}\ \bibnamefont
  {Kirkpatrick}},\ }\bibfield  {title} {\enquote {\bibinfo {title} {Possible
  energy gain for a plasma-liner-driven magneto-inertial fusion concept},}\
  }\href@noop {} {\bibfield  {journal} {\bibinfo  {journal} {Physics of
  Plasmas}\ }\textbf {\bibinfo {volume} {21}} (\bibinfo {year}
  {2014})}\BibitemShut {NoStop}%
\bibitem [{\citenamefont {Witherspoon}\ \emph {et~al.}(2009)\citenamefont
  {Witherspoon}, \citenamefont {Case}, \citenamefont {Messer}, \citenamefont
  {Bomgardner~II}, \citenamefont {Phillips}, \citenamefont {Brockington},\ and\
  \citenamefont {Eltonc}}]{myID_513}%
  \BibitemOpen
  \bibfield  {author} {\bibinfo {author} {\bibfnamefont {F.~D.}\ \bibnamefont
  {Witherspoon}}, \bibinfo {author} {\bibfnamefont {A.}~\bibnamefont {Case}},
  \bibinfo {author} {\bibfnamefont {S.~J.}\ \bibnamefont {Messer}}, \bibinfo
  {author} {\bibfnamefont {R.}~\bibnamefont {Bomgardner~II}}, \bibinfo {author}
  {\bibfnamefont {M.~W.}\ \bibnamefont {Phillips}}, \bibinfo {author}
  {\bibfnamefont {S.}~\bibnamefont {Brockington}},\ and\ \bibinfo {author}
  {\bibfnamefont {R.}~\bibnamefont {Eltonc}},\ }\bibfield  {title} {\enquote
  {\bibinfo {title} {A contoured gap coaxial plasma gun with injected plasma
  armature},}\ }\href {https://doi.org/10.1063/1.3202136} {\bibfield  {journal}
  {\bibinfo  {journal} {Review of Scientific Instruments}\ }\textbf {\bibinfo
  {volume} {80}} (\bibinfo {year} {2009}),\ 10.1063/1.3202136}\BibitemShut
  {NoStop}%
\bibitem [{\citenamefont {Merritt}\ \emph
  {et~al.}(2012{\natexlab{a}})\citenamefont {Merritt}, \citenamefont {Moser},
  \citenamefont {Hsu}, \citenamefont {Adams}, \citenamefont {Dunn},
  \citenamefont {Holgado},\ and\ \citenamefont {Gilmore}}]{myID_516}%
  \BibitemOpen
  \bibfield  {author} {\bibinfo {author} {\bibfnamefont {E.~C.}\ \bibnamefont
  {Merritt}}, \bibinfo {author} {\bibfnamefont {A.~L.}\ \bibnamefont {Moser}},
  \bibinfo {author} {\bibfnamefont {S.~C.}\ \bibnamefont {Hsu}}, \bibinfo
  {author} {\bibfnamefont {C.~S.}\ \bibnamefont {Adams}}, \bibinfo {author}
  {\bibfnamefont {J.~P.}\ \bibnamefont {Dunn}}, \bibinfo {author}
  {\bibfnamefont {A.~M.}\ \bibnamefont {Holgado}},\ and\ \bibinfo {author}
  {\bibfnamefont {M.~A.}\ \bibnamefont {Gilmore}},\ }\bibfield  {title}
  {\enquote {\bibinfo {title} {Experimental evidence for collisional shock
  formation via two obliquely merging supersonic plasma jets},}\ }\href
  {https://doi.org/10.1063/1.4872323} {\bibfield  {journal} {\bibinfo
  {journal} {Physics of Plasmas}\ }\textbf {\bibinfo {volume} {21}} (\bibinfo
  {year} {2012}{\natexlab{a}}),\ 10.1063/1.4872323}\BibitemShut {NoStop}%
\bibitem [{\citenamefont {Langendorf}\ \emph {et~al.}(2019)\citenamefont
  {Langendorf}, \citenamefont {Yates}, \citenamefont {Hsu}, \citenamefont
  {Thoma},\ and\ \citenamefont {Gilmore}}]{myID_570}%
  \BibitemOpen
  \bibfield  {author} {\bibinfo {author} {\bibfnamefont {S.~J.}\ \bibnamefont
  {Langendorf}}, \bibinfo {author} {\bibfnamefont {K.~C.}\ \bibnamefont
  {Yates}}, \bibinfo {author} {\bibfnamefont {S.~C.}\ \bibnamefont {Hsu}},
  \bibinfo {author} {\bibfnamefont {C.}~\bibnamefont {Thoma}},\ and\ \bibinfo
  {author} {\bibfnamefont {M.}~\bibnamefont {Gilmore}},\ }\bibfield  {title}
  {\enquote {\bibinfo {title} {Experimental study of ion heating in obliquely
  merging hypersonic plasma jets},}\ }\href {https://doi.org/doi:
  10.1063/1.5108727} {\bibfield  {journal} {\bibinfo  {journal} {Physics of
  Plasmas}\ }\textbf {\bibinfo {volume} {26}} (\bibinfo {year} {2019}),\ doi:
  10.1063/1.5108727}\BibitemShut {NoStop}%
\bibitem [{\citenamefont {Yates}\ \emph {et~al.}(2020)\citenamefont {Yates},
  \citenamefont {Langendorf}, \citenamefont {Hsu}, \citenamefont {Dunn},
  \citenamefont {Brockington}, \citenamefont {Case}, \citenamefont {Cruz},
  \citenamefont {Witherspoon}, \citenamefont {Thio}, \citenamefont {Cassibry},
  \citenamefont {Schillo},\ and\ \citenamefont {Gilmore}}]{myID_549}%
  \BibitemOpen
  \bibfield  {author} {\bibinfo {author} {\bibfnamefont {K.~C.}\ \bibnamefont
  {Yates}}, \bibinfo {author} {\bibfnamefont {S.~J.}\ \bibnamefont
  {Langendorf}}, \bibinfo {author} {\bibfnamefont {S.~C.}\ \bibnamefont {Hsu}},
  \bibinfo {author} {\bibfnamefont {J.~P.}\ \bibnamefont {Dunn}}, \bibinfo
  {author} {\bibfnamefont {S.}~\bibnamefont {Brockington}}, \bibinfo {author}
  {\bibfnamefont {A.}~\bibnamefont {Case}}, \bibinfo {author} {\bibfnamefont
  {E.}~\bibnamefont {Cruz}}, \bibinfo {author} {\bibfnamefont {F.~D.}\
  \bibnamefont {Witherspoon}}, \bibinfo {author} {\bibfnamefont {Y.~C.~F.}\
  \bibnamefont {Thio}}, \bibinfo {author} {\bibfnamefont {J.~T.}\ \bibnamefont
  {Cassibry}}, \bibinfo {author} {\bibfnamefont {K.}~\bibnamefont {Schillo}},\
  and\ \bibinfo {author} {\bibfnamefont {M.}~\bibnamefont {Gilmore}},\
  }\bibfield  {title} {\enquote {\bibinfo {title} {Experimental
  characterization of a section of a spherically imploding plasma liner formed
  by merging hypersonic plasma jets},}\ }\href
  {https://doi.org/10.1063/1.5126855} {\bibfield  {journal} {\bibinfo
  {journal} {Physics of Plasmas}\ }\textbf {\bibinfo {volume} {27}} (\bibinfo
  {year} {2020}),\ 10.1063/1.5126855}\BibitemShut {NoStop}%
\bibitem [{\citenamefont {Langendorf}\ \emph {et~al.}(2018)\citenamefont
  {Langendorf}, \citenamefont {Yates}, \citenamefont {Hsu}, \citenamefont
  {Thoma},\ and\ \citenamefont {Gilmore}}]{myID_512}%
  \BibitemOpen
  \bibfield  {author} {\bibinfo {author} {\bibfnamefont {S.~J.}\ \bibnamefont
  {Langendorf}}, \bibinfo {author} {\bibfnamefont {K.~C.}\ \bibnamefont
  {Yates}}, \bibinfo {author} {\bibfnamefont {S.~C.}\ \bibnamefont {Hsu}},
  \bibinfo {author} {\bibfnamefont {C.}~\bibnamefont {Thoma}},\ and\ \bibinfo
  {author} {\bibfnamefont {M.}~\bibnamefont {Gilmore}},\ }\bibfield  {title}
  {\enquote {\bibinfo {title} {Experimental measurements of ion heating in
  collisional plasma shocks and interpenetrating supersonic plasma flows},}\
  }\href {https://doi.org/10.1103/PhysRevLett.121.185001} {\bibfield  {journal}
  {\bibinfo  {journal} {Physical Review Letters}\ }\textbf {\bibinfo {volume}
  {121}} (\bibinfo {year} {2018}),\ 10.1103/PhysRevLett.121.185001}\BibitemShut
  {NoStop}%
\bibitem [{\citenamefont {Hsu}\ \emph {et~al.}(2015)\citenamefont {Hsu},
  \citenamefont {Moser}, \citenamefont {Merritt}, \citenamefont {Adams},
  \citenamefont {Dunn}, \citenamefont {Brockington}, \citenamefont {Case},
  \citenamefont {Gilmore}, \citenamefont {Lynn}, \citenamefont {Messer} \emph
  {et~al.}}]{myID_576}%
  \BibitemOpen
  \bibfield  {author} {\bibinfo {author} {\bibfnamefont {S.~C.}\ \bibnamefont
  {Hsu}}, \bibinfo {author} {\bibfnamefont {A.~L.}\ \bibnamefont {Moser}},
  \bibinfo {author} {\bibfnamefont {E.~C.}\ \bibnamefont {Merritt}}, \bibinfo
  {author} {\bibfnamefont {C.~S.}\ \bibnamefont {Adams}}, \bibinfo {author}
  {\bibfnamefont {J.~P.}\ \bibnamefont {Dunn}}, \bibinfo {author}
  {\bibfnamefont {S.}~\bibnamefont {Brockington}}, \bibinfo {author}
  {\bibfnamefont {A.}~\bibnamefont {Case}}, \bibinfo {author} {\bibfnamefont
  {M.}~\bibnamefont {Gilmore}}, \bibinfo {author} {\bibfnamefont {A.~G.}\
  \bibnamefont {Lynn}}, \bibinfo {author} {\bibfnamefont {S.~J.}\ \bibnamefont
  {Messer}}, \emph {et~al.},\ }\bibfield  {title} {\enquote {\bibinfo {title}
  {Laboratory plasma physics experiments using merging supersonic plasma
  jets},}\ }\href@noop {} {\bibfield  {journal} {\bibinfo  {journal} {Journal
  of Plasma Physics}\ }\textbf {\bibinfo {volume} {81}},\ \bibinfo {pages}
  {345810201} (\bibinfo {year} {2015})}\BibitemShut {NoStop}%
\bibitem [{\citenamefont {Hsu}\ \emph {et~al.}(2012{\natexlab{b}})\citenamefont
  {Hsu}, \citenamefont {Merritt}, \citenamefont {Moser}, \citenamefont {Awe},
  \citenamefont {Brockington}, \citenamefont {Davis}, \citenamefont {Adams},
  \citenamefont {Case}, \citenamefont {Cassibry}, \citenamefont {Dunn},
  \citenamefont {Gilmore}, \citenamefont {Lynn}, \citenamefont {Messer},\ and\
  \citenamefont {Witherspoon}}]{myID_580}%
  \BibitemOpen
  \bibfield  {author} {\bibinfo {author} {\bibfnamefont {S.~C.}\ \bibnamefont
  {Hsu}}, \bibinfo {author} {\bibfnamefont {E.~C.}\ \bibnamefont {Merritt}},
  \bibinfo {author} {\bibfnamefont {A.~L.}\ \bibnamefont {Moser}}, \bibinfo
  {author} {\bibfnamefont {T.~J.}\ \bibnamefont {Awe}}, \bibinfo {author}
  {\bibfnamefont {S.~J.~E.}\ \bibnamefont {Brockington}}, \bibinfo {author}
  {\bibfnamefont {J.~S.}\ \bibnamefont {Davis}}, \bibinfo {author}
  {\bibfnamefont {C.~S.}\ \bibnamefont {Adams}}, \bibinfo {author}
  {\bibfnamefont {A.}~\bibnamefont {Case}}, \bibinfo {author} {\bibfnamefont
  {J.~T.}\ \bibnamefont {Cassibry}}, \bibinfo {author} {\bibfnamefont {J.~P.}\
  \bibnamefont {Dunn}}, \bibinfo {author} {\bibfnamefont {M.~A.}\ \bibnamefont
  {Gilmore}}, \bibinfo {author} {\bibfnamefont {A.~G.}\ \bibnamefont {Lynn}},
  \bibinfo {author} {\bibfnamefont {S.~J.}\ \bibnamefont {Messer}},\ and\
  \bibinfo {author} {\bibfnamefont {F.~D.}\ \bibnamefont {Witherspoon}},\
  }\bibfield  {title} {\enquote {\bibinfo {title} {{Experimental
  characterization of railgun-driven supersonic plasma jets motivated by high
  energy density physics applications}},}\ }\href
  {https://doi.org/10.1063/1.4773320} {\bibfield  {journal} {\bibinfo
  {journal} {Physics of Plasmas}\ }\textbf {\bibinfo {volume} {19}},\ \bibinfo
  {pages} {123514} (\bibinfo {year} {2012}{\natexlab{b}})},\ \Eprint
  {https://arxiv.org/abs/https://pubs.aip.org/aip/pop/article-pdf/doi/10.1063/1.4773320/14087096/123514\_1\_online.pdf}
  {https://pubs.aip.org/aip/pop/article-pdf/doi/10.1063/1.4773320/14087096/123514\_1\_online.pdf}
  \BibitemShut {NoStop}%
\bibitem [{\citenamefont {LaJoie}\ \emph {et~al.}(2023)\citenamefont {LaJoie},
  \citenamefont {Chu}, \citenamefont {Langendorf}, \citenamefont {Cassibry},
  \citenamefont {Vyas},\ and\ \citenamefont {Gilmore}}]{myID_556}%
  \BibitemOpen
  \bibfield  {author} {\bibinfo {author} {\bibfnamefont {A.~L.}\ \bibnamefont
  {LaJoie}}, \bibinfo {author} {\bibfnamefont {F.}~\bibnamefont {Chu}},
  \bibinfo {author} {\bibfnamefont {S.}~\bibnamefont {Langendorf}}, \bibinfo
  {author} {\bibfnamefont {J.}~\bibnamefont {Cassibry}}, \bibinfo {author}
  {\bibfnamefont {A.}~\bibnamefont {Vyas}},\ and\ \bibinfo {author}
  {\bibfnamefont {M.}~\bibnamefont {Gilmore}},\ }\bibfield  {title} {\enquote
  {\bibinfo {title} {Multi-camera imaging to characterize jet and liner
  uniformity on the plasma liner experiment (plx)},}\ }\href
  {https://doi.org/10.1063/5.0101674} {\bibfield  {journal} {\bibinfo
  {journal} {Review of Scientific Instruments}\ }\textbf {\bibinfo {volume}
  {94}} (\bibinfo {year} {2023}),\ 10.1063/5.0101674}\BibitemShut {NoStop}%
\bibitem [{\citenamefont {Merritt}\ \emph
  {et~al.}(2012{\natexlab{b}})\citenamefont {Merritt}, \citenamefont {Lynn},
  \citenamefont {Gilmore},\ and\ \citenamefont {Hsu}}]{myID_517}%
  \BibitemOpen
  \bibfield  {author} {\bibinfo {author} {\bibfnamefont {E.~C.}\ \bibnamefont
  {Merritt}}, \bibinfo {author} {\bibfnamefont {A.~G.}\ \bibnamefont {Lynn}},
  \bibinfo {author} {\bibfnamefont {M.~A.}\ \bibnamefont {Gilmore}},\ and\
  \bibinfo {author} {\bibfnamefont {S.~C.}\ \bibnamefont {Hsu}},\ }\bibfield
  {title} {\enquote {\bibinfo {title} {Multi-chord fiber-coupled interferometer
  with a long coherence length laser},}\ }\href
  {https://doi.org/10.1063/1.3697731} {\bibfield  {journal} {\bibinfo
  {journal} {Review of Scientific Instruments}\ }\textbf {\bibinfo {volume}
  {83}} (\bibinfo {year} {2012}{\natexlab{b}}),\ 10.1063/1.3697731}\BibitemShut
  {NoStop}%
\bibitem [{\citenamefont {Chu}\ \emph {et~al.}(2023)\citenamefont {Chu},
  \citenamefont {LaJoie}, \citenamefont {Keenan}, \citenamefont {Webster},
  \citenamefont {Langendorf},\ and\ \citenamefont {Gilmore}}]{myID_555}%
  \BibitemOpen
  \bibfield  {author} {\bibinfo {author} {\bibfnamefont {F.}~\bibnamefont
  {Chu}}, \bibinfo {author} {\bibfnamefont {A.~L.}\ \bibnamefont {LaJoie}},
  \bibinfo {author} {\bibfnamefont {B.~D.}\ \bibnamefont {Keenan}}, \bibinfo
  {author} {\bibfnamefont {L.}~\bibnamefont {Webster}}, \bibinfo {author}
  {\bibfnamefont {S.~J.}\ \bibnamefont {Langendorf}},\ and\ \bibinfo {author}
  {\bibfnamefont {M.}~\bibnamefont {Gilmore}},\ }\bibfield  {title} {\enquote
  {\bibinfo {title} {Experimental measurements of ion diffusion coefficients
  and heating in a multi-ion-species plasma shock},}\ }\href
  {https://doi.org/10.1103/PhysRevLett.130.145101} {\bibfield  {journal}
  {\bibinfo  {journal} {Physical Review Letters}\ }\textbf {\bibinfo {volume}
  {130}} (\bibinfo {year} {2023}),\ 10.1103/PhysRevLett.130.145101}\BibitemShut
  {NoStop}%
\bibitem [{\citenamefont {Prince}\ and\ \citenamefont
  {Robertson}(1966)}]{myID_561}%
  \BibitemOpen
  \bibfield  {author} {\bibinfo {author} {\bibfnamefont {J.~F.}\ \bibnamefont
  {Prince}}\ and\ \bibinfo {author} {\bibfnamefont {W.~W.}\ \bibnamefont
  {Robertson}},\ }\bibfield  {title} {\enquote {\bibinfo {title} {Continuum
  radiation in an argon positive column},}\ }\href
  {https://doi.org/10.1063/1.1727977} {\bibfield  {journal} {\bibinfo
  {journal} {Journal of Chemical Physics}\ }\textbf {\bibinfo {volume} {45}},\
  \bibinfo {pages} {2577–2584} (\bibinfo {year} {1966})}\BibitemShut
  {NoStop}%
\bibitem [{\citenamefont {Wilbers}\ \emph {et~al.}(1991)\citenamefont
  {Wilbers}, \citenamefont {Kroesen}, \citenamefont {Timmermans},\ and\
  \citenamefont {Schram}}]{myID_562}%
  \BibitemOpen
  \bibfield  {author} {\bibinfo {author} {\bibfnamefont {A.}~\bibnamefont
  {Wilbers}}, \bibinfo {author} {\bibfnamefont {G.}~\bibnamefont {Kroesen}},
  \bibinfo {author} {\bibfnamefont {C.}~\bibnamefont {Timmermans}},\ and\
  \bibinfo {author} {\bibfnamefont {D.}~\bibnamefont {Schram}},\ }\bibfield
  {title} {\enquote {\bibinfo {title} {The continuum emission of an arc
  plasma},}\ }\href {https://doi.org/10.1016/0022-4073(91)90076-3} {\bibfield
  {journal} {\bibinfo  {journal} {Journal of Quantitative Spectroscopy and
  Radiative Transfer}\ }\textbf {\bibinfo {volume} {45}},\ \bibinfo {pages}
  {1--10} (\bibinfo {year} {1991})}\BibitemShut {NoStop}%
\bibitem [{\citenamefont {Aparicio}\ \emph {et~al.}(1998)\citenamefont
  {Aparicio}, \citenamefont {Gigosos}, \citenamefont {Gonz\'{a}lez},
  \citenamefont {P\'{e}rez}, \citenamefont {de~la Rosa},\ and\ \citenamefont
  {Mar}}]{myID_560}%
  \BibitemOpen
  \bibfield  {author} {\bibinfo {author} {\bibfnamefont {J.~A.}\ \bibnamefont
  {Aparicio}}, \bibinfo {author} {\bibfnamefont {M.~A.}\ \bibnamefont
  {Gigosos}}, \bibinfo {author} {\bibfnamefont {V.~R.}\ \bibnamefont
  {Gonz\'{a}lez}}, \bibinfo {author} {\bibfnamefont {C.}~\bibnamefont
  {P\'{e}rez}}, \bibinfo {author} {\bibfnamefont {M.~I.}\ \bibnamefont {de~la
  Rosa}},\ and\ \bibinfo {author} {\bibfnamefont {S.}~\bibnamefont {Mar}},\
  }\bibfield  {title} {\enquote {\bibinfo {title} {Measurement of stark
  broadening and shift of singly ionized ar lines},}\ }\href
  {https://doi.org/10.1088/0953-4075/31/5/011} {\bibfield  {journal} {\bibinfo
  {journal} {Journal of Physics B: Atomic, Molecular and Optical Physics}\
  }\textbf {\bibinfo {volume} {31}},\ \bibinfo {pages} {1029--1048} (\bibinfo
  {year} {1998})}\BibitemShut {NoStop}%
\bibitem [{\citenamefont {Schillo}\ and\ \citenamefont
  {Cassibry}(2020)}]{myID_581}%
  \BibitemOpen
  \bibfield  {author} {\bibinfo {author} {\bibfnamefont {K.}~\bibnamefont
  {Schillo}}\ and\ \bibinfo {author} {\bibfnamefont {J.}~\bibnamefont
  {Cassibry}},\ }\bibfield  {title} {\enquote {\bibinfo {title} {Effects of
  initial conditions and transport on ram pressure, mach number, and uniformity
  for plasma liner formation and implosion},}\ }\href@noop {} {\bibfield
  {journal} {\bibinfo  {journal} {Physics of Plasmas}\ }\textbf {\bibinfo
  {volume} {27}} (\bibinfo {year} {2020})}\BibitemShut {NoStop}%
\bibitem [{\citenamefont {Rolison}\ \emph {et~al.}(2019)\citenamefont
  {Rolison}, \citenamefont {Fensin}, \citenamefont {Thio}, \citenamefont
  {Hsu},\ and\ \citenamefont {Cruz}}]{myID_571}%
  \BibitemOpen
  \bibfield  {author} {\bibinfo {author} {\bibfnamefont {L.~M.}\ \bibnamefont
  {Rolison}}, \bibinfo {author} {\bibfnamefont {M.~L.}\ \bibnamefont {Fensin}},
  \bibinfo {author} {\bibfnamefont {Y.~C.~F.}\ \bibnamefont {Thio}}, \bibinfo
  {author} {\bibfnamefont {S.~C.}\ \bibnamefont {Hsu}},\ and\ \bibinfo {author}
  {\bibfnamefont {E.~J.}\ \bibnamefont {Cruz}},\ }\bibfield  {title} {\enquote
  {\bibinfo {title} {Neutronics calculations for a hypothetical
  plasma-jet-driven magneto-inertial-fusion reactor},}\ }\href@noop {}
  {\bibfield  {journal} {\bibinfo  {journal} {Fusion Science and Technology}\
  }\textbf {\bibinfo {volume} {75}},\ \bibinfo {pages} {438--451} (\bibinfo
  {year} {2019})}\BibitemShut {NoStop}%
\end{thebibliography}%

\end{document}